\documentclass[10pt,twoside]{classe-cmb}
\usepackage{amssymb,amsbsy,amsmath,amsfonts,amssymb,amscd}
\usepackage{latexsym,euscript,exscale,epic,eepic,epsfig}
\usepackage[english,francais]{babel}
\usepackage{times}
\usepackage{color}

\newcommand{\dd}{{\rm d}}
\newcommand{\etal}{{\em et al.}}
\ComParit{S1296-2147}
\PIT{FLA}
\PXHY{????}
\Add{?}
\Volume{0} \Year{2014} \FirstPage{1} \LastPage{??}
\AuteurCourant{
G.F.R. Ellis and J.-P. Uzan
}
\TitreCourant{Causal structures in cosmology} 
\Journal
\Rubrique{Rubrique}{Heading}
\SousRubrique{Sous-rubrique}{Sub-Heading}
\PresentePar{Jean-Philippe}{Uzan}
 \Recu{blank}{}{}
\begin{document}
\selectlanguage{english}
\TitleOfDossier{Inflation: theoretical and observational status}
\title{%
 Causal structures in inflation
 \vskip0.2cm
 Structures causales de l'inflation}
\author{%
George F.R. Ellis~$^{\text{a}}$,\ \ Jean-Philippe Uzan$^{\text{b,c}}$}
\address{%
\begin{itemize}\labelsep=2mm\leftskip=-5mm
\item[$^{\text{a}}$] Mathematics and Applied Mathematics Department,
University of Cape Town, Rondebosch, Cape Town 7701, South Africa,\\
E-mail: george.ellis@uct.ac.za
\item[$^{\text{b}}$] Institut d'Atrophysique de Paris, UMR-7095 du CNRS, Universit\'e Pierre et Marie
              Curie,\\
98 bis, boulevard Arago, 75014 Paris,
France,\\
\item[$^{\text{c}}$]  Sorbonne Universit\'es, Institut Lagrange de Paris,\\
98 bis, boulevard Arago, 75014 Paris, France.\\
E-mail: uzan@iap.fr
\end{itemize}
}
\maketitle
\thispagestyle{empty}
\begin{Abstract}
{This article reviews the properties and limitations associated with the existence of particle, visual, and event horizons in cosmology in general and in inflationary universes in particular, carefully distinguishing them from `Hubble horizons'. It explores to what extent one might be able to probe conditions beyond the visual horizon (which is close in size to the present Hubble radius), thereby showing that visual horizons place major limits on what are observationally testable aspects of a multiverse, if such exists. Indeed these limits largely prevent us from observationally proving a multiverse either does or does not exist. We emphasize that event horizons play no role at all in observational cosmology, even in the multiverse context, despite some claims to the contrary in the literature.\vskip.25cm
\hskip2cm Cet article d\?efinit puis passe en revue les propri\?et\?es et les limites associ\?ees \`a l'existence d'horizons (des particules, des \?ev\?enements et visuels) en cosmologie en g\?en\?eral et pendant l'inflation en particulier, en insistant sur leurs diff\?erences avec les `horizons de Hubble'. Il discute la possibilit\?e de tester les conditions physiques de l'univers au-del\`a de notre horizon visuel (qui, en taille, est proche du rayon de Hubble) et démontre que l'existence d'horizons visuels impose des limites strictes sur ce qui est potentiellement testable dans les sc\'enarios de type multivers, si ils existent. Ces limites nous interdisent de prouver observationnellement l'existence ou la non-existence de ces multivers. Il est aussi d\'emontr\'e que les horizons des \?ev\?enements ne jouent aucun r\^ole en cosmologie observationnelle, même dans le contexte des mod\`eles de multivers.}
\end{Abstract}

\par\medskip\centerline{\rule{2cm}{0.2mm}}\medskip
\setcounter{section}{0}
\selectlanguage{english}
\twocolumn

\section{Introduction}

The causal structure of spacetimes plays a major role in the understanding of the physics of black holes and in cosmology. In particular these spacetime possess horizons. A horizon is a frontier that bounds causality, or  separates observable events from non-observable ones. In cosmology they limit the observational possibilities, and they have to be distinguished from the natural scales fixed by the cosmic expansion rate. The way they do so differs in non-inflationary 
and inflationary cosmology. {\color{blue}}

These hypersurfaces play different roles in their two main contexts, the physics of black holes and cosmology~\cite{Pen64,Rin56,HawEll73,TipClaEll80,PetUza13,DerUza14}. We focus here on their role in cosmology, but contrast this with the black hole case. There appears to be substantial confusion about this in some of the current literature on inflationary cosmology, where in particular event horizons are claimed to play a significant physical role; but this is not the case.  

The article is organized as follows: Section~\ref{sec1} defines the different notions of horizons, and Section~\ref{sec1b} focuses on the use of conformal diagrams.  Section~\ref{sec3} discusses the case of standard cosmology, while Sections~\ref{sec4} and~\ref{sec6} consider inflationary cosmology and alternative models, including small universes and so\-me multiverse proposals.

\section{Different notions of horizons}\label{sec1}

Very different concepts of horizons have to be considered. In particular, one needs to distinguish between {\em local} and {\em non-local} (or {\em global})  notions of horizons, respectively defined in \S~\ref{sec22} and \S~\ref{sec23}.

In order to introduce all these notions, we firstly assume that the spacetimes under consideration are globally hyperbolic so that they can be foliated by a continuous family of spacelike three-dimensional hypersurfaces, $\Sigma_t$ \cite{HawEll73}. This means that there exists a smooth function $\hat t$ on ${\cal M}$ whose gradient never vanishes and is timelike so that each hypersurface is a surface of constant $\hat t$,
\begin{equation}
\Sigma_t=\lbrace p\in{\cal M},\, \hat t(p)=t\rbrace,
\end{equation}
$\forall t\in I \subset \mathbb{R}$ and $g^{\mu\nu} \hat t_\mu \hat t_\nu<0$ where $I$ is a maximal subset of $\mathbb{R}$ so that $\Sigma_t$ covers all ${\cal M}$.  Such spacetimes represent most spacetimes of astrophysical and cosmological interest. They imply existence of a global direction of time. The expansion of the universe, and associated physically meaningful horizons, occur relative to the future direction of time. 

Secondly, in the cosmological case, we assume existence everywhere of a family of fundamental observers with 4-velocity $u_\mu: u^\mu u_\mu = -1$, defining a preferred cosmological restframe at each point \cite{ehl61,ell71}. This implies that the worldlines of fundamental observers never intersect.

\subsection{Past Light Cone}\label{sec21}

Given a spacetime ${\cal M}$ with metric $g_{\mu\nu}$, one can define for any event $p$ the past lightcone $C^-(p)$ as the set of events $q$ such that there exists a future directed null geodesic joining $q$ to $p$. It characterizes the set of events that can be observed by an observer at event $p$ by electromagnetic radiation, irrespective of its wavelength. Technically, for any event $q$ on $C^-(p)$, there exists a null geodesic $x^\mu(\lambda)$ parameterized by the affine parameter $\lambda\leq0$ (chosen negative so that increasing $\lambda$ corresponds to the future direction of time), such that $x^\mu(0) = p$ and there is a value $\lambda_1$ such that $x^\mu(\lambda_1) = q$. Its tangent vector $k^\mu\equiv d x^\mu/d\lambda$ satisfies the null geodesic equation
\begin{equation}\label{e.null}
 k^\mu k_\mu=0,
 \qquad
 k^\mu\nabla_\mu k^\nu=0.
\end{equation}
The past lightcone is a 3-dimensional null surface that can be parameterized by 2 angles ($\theta, \phi$) representing the direction of observation in the sky and a redshift $z$ that characterizes the distance ``down'' the lightcone, and is defined as
\begin{equation}
1+z\equiv \frac{(-k^\mu u_\mu)_{\rm source}}{(-k^\mu u_\mu)_{\rm obs}}
\end{equation}
where $u^\mu$ is the tangent vector to the observer and source worldlines.  These quantities are defined with respect to the fundamental observers. By construction, $C^-(p)$ depends on the event $p$ so that two different observers have different lightcones, and any specific observer's lightcone changes over time. In cosmology, this latter effect is at the origin of the time drift of observed redshift~\cite{zdot}.  

Let us recall an important property. Consider two spacetimes whose metrics $g_{\mu\nu}$ and $\tilde g_{\mu\nu}$ are conformal, i.e. $g_{\mu\nu}=\Omega^2 \tilde g_{\mu\nu}$. Any null geodesic of $g_{\mu\nu}$ with affine parameter $\lambda$ is a null geodesic of  $\tilde g_{\mu\nu}$ with affine parameter $\tilde \lambda$ where $d\lambda=\Omega^2d\tilde\lambda$, so that $\tilde k^\mu=\Omega^2 k^\mu$; see e.g. Ref.~\cite{wald}.

\subsection{Non-local Horizons}\label{sec22}

We can define several types of non-local horizons, namely particle horizons \cite{Rin56}, visual horizons~\cite{EllSto88,EllRot93} and event horizons~\cite{Rin56}. They are non-local in the sense that they depend on the large-scale geodesic structure of the spacetime. They are defined relative to the future direction of time. 

\vskip0.25cm
\noindent[A] {\it\bf Particle Horizons (PH)}.

Particle horizons are defined as the worldlines of the limiting fundamental particles that can affect an observer $O$ at a spacetime position $p$~ \cite{Rin56}, provided such a limit exists.  For an observer $O$ at time $t_0$, the particle horizon is the timelike hypersurface that at time $t = t_0$ divides all particles in the Universe into two non-empty families: the ones that can have already been observed or been in causal contact with $O$ at the time $t_0$, and the ones that cannot  have been observed or in causal contact then. For each time $t_0$, the 
particle horizon is determined as  the intersection between the limiting geodesics of the most distant comoving particles that can be causally interacted with (they lie on the past light cone of $C^-(p)$ of $p$), with the past hypersurface $\Sigma_{t{_1}}$ as $t_1$ is taken to the limits of the boundary of the spacetime ($t_1$ is not necessary finite). This limit is the limit of a two-dimensional spacelike surface which reduces to a sphere of centre $O$ if the spacetime enjoys a rotational symmetry around the observers worldline. Clearly this depends on the foliation of spacetime by surfaces $\{t={\rm const}\}$; in a spatially homogeneous Friedmann-Lema\^{\i}tre (FL) geometry, the natural such foliation is unique.

In cosmology, the PH for an observer at the present time $t_0$ is a key concept in discussing causal limits of an observer at the present time. It relates to the start of the universe, or initial boundary of the universe if there is no beginning. One can of course also define particle horizons for events at earlier times than the present, for example the particle horizon at the time of  recombination, where it will characterize causal limits for events at that time (this is the Primordial Particle Horizon, see Section 4.2[C]).   Note also that if there are closed spatial sections, perhaps due to  a non-trivial (i.e., non-simply connected) topology (see below), the spacetime may not have any  particle horizons at late enough times, as is also the case if  its expansion is accelerated with a non-singular start.

\vskip0.25cm
\noindent[B] {\it\bf Visual horizons (VH)}.

\begin{figure}[t!]
\begin{center}
\resizebox{8cm}{!}{\includegraphics[clip=true]{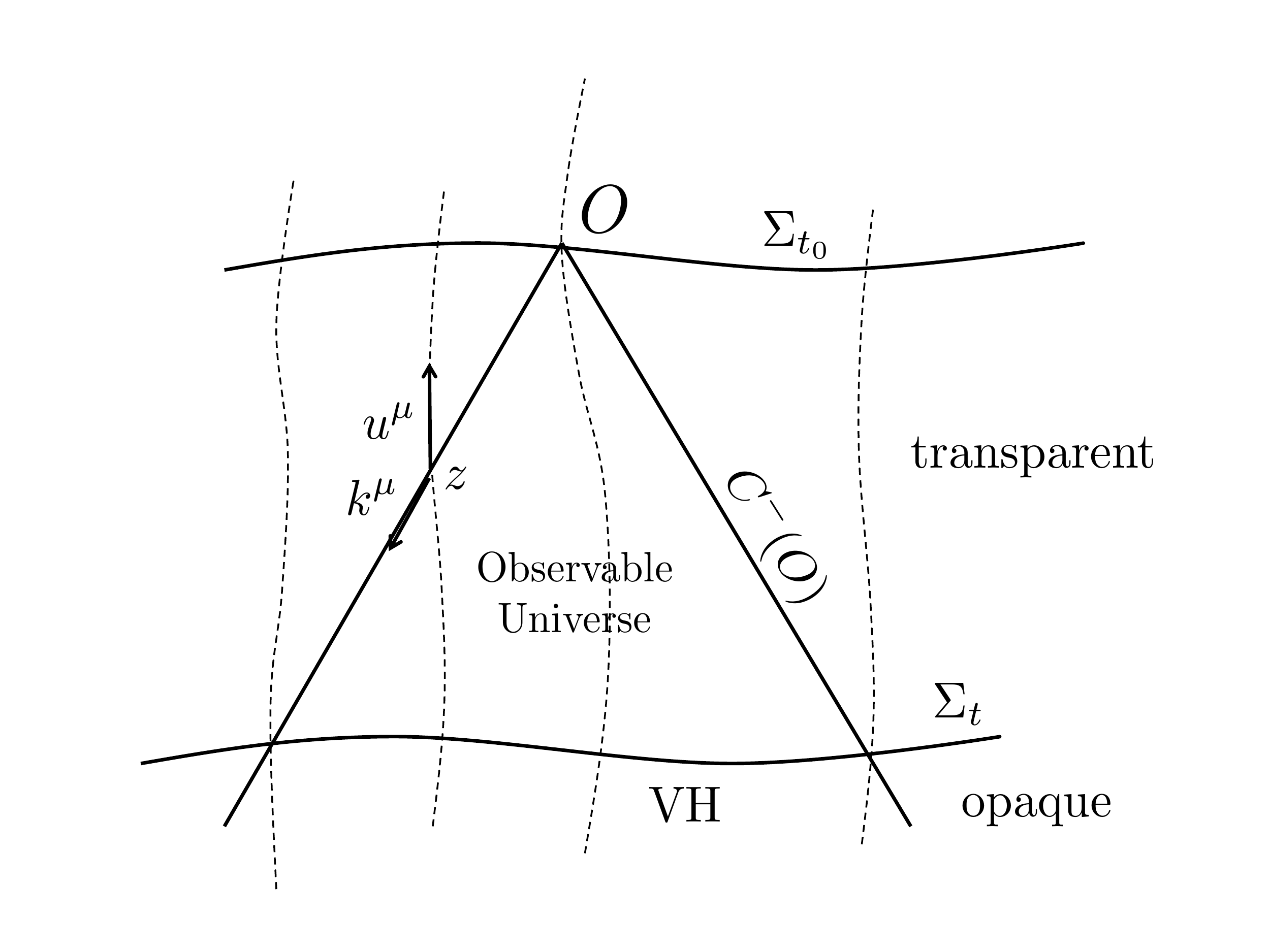}}
\end{center}
\vskip-0.5cm
 \caption{\footnotesize The past lightcone, visual horizon and observable universe of an observer $O$ in a globally hyperbolic spacetime.}
  \label{fig2}
\end{figure}

If the past lightcone intersects a spatial section $\Sigma_{t_d}$ before which the spacetime is opaque to electromagnetic radiation, this leads to the existence of a visual horizon for electromagnetic radiation, defined as the set of fundamental worldlines passing through $\Sigma_p := C^-(p)\cap \Sigma_{t_d}$. We cannot see further out matter by any form of electromagnetic radiation, and the observable universe for $p$ is the spacetime region delimited by the past light cone $C^-(p)$ up to $\Sigma_p$ (see Fig.~\ref{fig2}). A typical example of such a visual horizon is the last scattering surface in cosmology which characterizes how far we can see by electromagnetic radiation~ \cite{EllSto88,EllRot93} so that it is a key concept in the discussion of our observational limits (see below). The 2-sphere $\Sigma_p$ delineates the furthest observable matter in the universe, for $p$; it is what we observe by means of cosmic background radiation observations, such as by COBE, Planck, and Bicep2. At later times $p'$, the corresponding horizon will move out and $\Sigma_{p'}$ will lie outside $\Sigma_p$. Visual horizons can also appear in black hole physics, when it is surrounded by an accretion disk.

It follows from the definition that the visual horizon contains the worldlines of all matter 
that we can observe by photons of any wavelength. 
By construction, all these worldlines lie inside the particle horizon. Let us also stress that we may have different visual horizons according to the messenger used to observe, respectively \textsc{VH}$(\gamma)$, \textsc{VH}$(\nu)$,  \textsc{VH({\small GW})} for photons, neutrinos, and gravity waves\footnote{Why not for cosmic rays also? The basic problem here is that because of magnetic fields they don't travel on geodesics in spacetime; hence their observed direction of arrival does not tell us where they came from, so  they are not good imaging tools.} respectively  since (1) they may not define the same ``cones'' ($\nu$ are supposed to be massive so that they propagate inside the lightcone contrary to photons and gravitational waves) and (2) the universe may not become opaque in the same circumstances (last scattering takes place earlier for neutrinos than for photons, and the universe remains transparent to gravity waves back to the Planck time).

Note also that if the spatial sections have a non-trivial topology, the spacetime may not have any visual horizons if the size of its fundamental domain is smaller than the size of its visual horizon in its covering space, see e.g. Refs.~\cite{topology,topogen}. Light can then travel right round the universe since $\Sigma_{t_d}$.

\vskip0.25cm
\noindent[C] {\it\bf Event horizons (EH)}.

For an observer $O$, the event horizon is the null  hypersurface that divides all events into two families: the ones that have been, are, or will ever be observable by $O$ or in causal contact with $O$, and the ones that are for ever outside the observational  and causal perimeter of $O$. Hence, it limits what can ever affect an observer in its entire history. Consequently it does not relate to any observations that can be made made at the present day \cite{Rin56}. It is a null surface that can be seen as the future lightcone of the observer's worldline in the limit $t\rightarrow+\infty$ (or $t \rightarrow t_+$ if there is a future singularity at $t = t_+$). Unlike the particle horizon, it is not defined for one instant in an observer's history: it depends on the entire future history of the observer, as well as the global spacetime structure.  
 
For black holes, EH are closely related to singularity existence, basically because they are related to apparent horizons (see below). In cosmology, it is important to realize that they have no relation whatsoever with observations or causal limits for us  at the present time, because they relate only to limits in the far distant future (see Section 6.3).\\

\begin{figure}[t!]
\begin{center}
\resizebox{6cm}{!}{\includegraphics[clip=true]{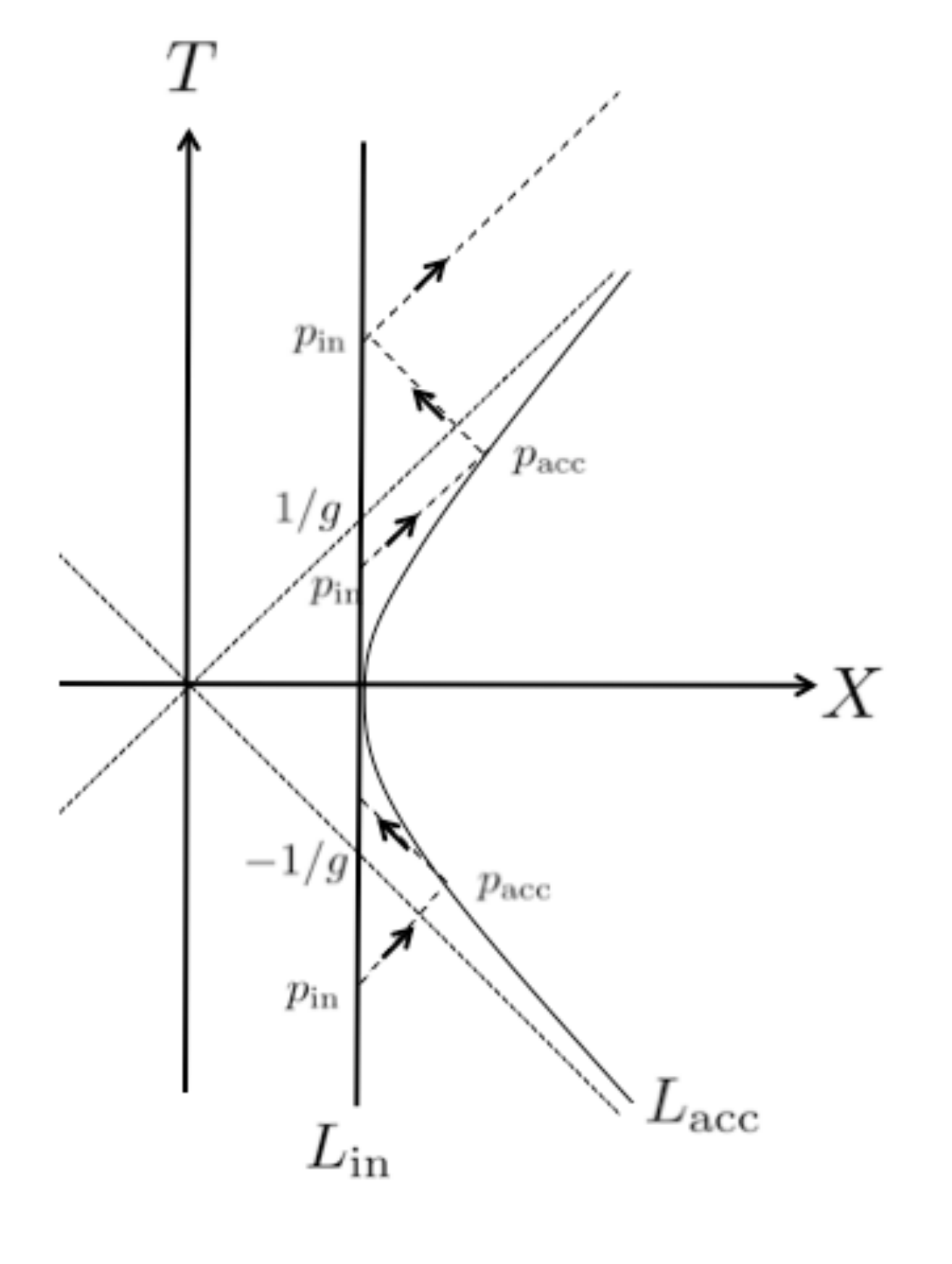}}
\end{center}
\vskip-0.5cm
 \caption{\footnotesize Worldlines of $P_{\rm in}$, at rest, and $P_{\rm acc}$, uniformly accelerated, in a Minkowski spacetime. If the aceleration is eternal, there exist event horizons: 
 $P_{\rm in}$ exits the event horizon of $P_{\rm acc}$ at $1/g$. There are no event horizons for $P_{in}$. From Ref.~\cite{DerUza14}.}
\label{fig1}
\end{figure}

Let us emphasize that such horizons can also appear in special relativity. Consider two observers, an inertial observer $P_{\rm in}$ at rest in a Minkowkian restframe such that its worldline is $X=1/g$, and $P_{\rm acc}(g)$ subject to a constant acceleration so that its worldline is given by $(X,T)=(\sinh g\tau,\cosh g\tau)/g$ with $g$ a constant; see Fig.~\ref{fig1}. Assume that $P_{\rm in}$ sends a signal at $T=T_e$. It propagates as $X=T-T_e+1/g$. Since the  worldline of $P_{\rm acc}$ enjoys $X=T$ as an asymptote when $T\rightarrow+\infty$, each worldline in the family of observers $P_{\rm acc}(g)$ for any $g$ has the past null surface $T = +X$ as an event horizon.   It is clear that no signal emitted by $P_{\rm in}$ after $T_+=1/g$ can reach $P_{\rm acc}$ (note that this requires that $P_{\rm acc}$ is accelerated {\it forever}).  This already illustrates the fact that event horizon are global quantities that depend on the whole structure of spacetime (or here the whole trajectory of observers that are under consideration). Hence $P_{\rm in}$ exits the event horizon of $P_{\rm acc}$ at $T_+$. We refer to Ref.~\cite{DerUza14} (p. 211) for the computation of the redshift of the signal. This property is important in the study of the Unruh effect.

\subsection{Local Horizons}\label{sec23}

Two notions of local horizons, which both are actually {\em not} horizons, are often considered in the literature:  the ``{\em Hubble Horizon}'', and apparent horizons. In contrast to particle, event, and visual horizons, their existence and location are determined by local inequalities. They are again defined relative to the future direction of time. 
 
\vskip0.25cm
\noindent[A] {\it\bf Hubble Horizon (HH)}. 

For a universe filled with a cosmic fluid, that is matter averaged to a cosmological scale where it can be regarded as a continuous fluid representing the average motion of all matter in the universe, one can define a normalised tangent vector $u^\mu$ to the corresponding fluid worldlines ($u^\mu u_\mu=-1$) \cite{ehl61,ell71}.  The local spatial metric is  then $h_{\mu\nu}:= g_{\mu\nu} +u_\mu u_\nu$. The relative motion of the fluid particles - which is what is observable in astronomical observations, when the `particles' are groups of galaxies  - is then characterised by a set of kinematic quantities \cite{HawEll73,ehl61,ell71}:
\begin{equation}
 \nabla_\mu u_\nu = \frac{1}{3}\Theta h_{\mu\nu}  + \sigma_{\mu\nu}+ \omega_{\mu\nu} - u_\mu u^\alpha\nabla_\alpha u_\nu,
\end{equation}
where the scalar $\Theta\equiv \nabla_\alpha u^\alpha$ is the rate of volume expansion of the fluid, $\sigma_{\mu\nu}$ is the trace-free symmetric shear tensor describing the rate of distortion of the matter flow, $\omega_{\mu\nu}$ is the skew symmetric vorticity tensor describing the rotation of the matter, and the last term accounts for non-geodesic motion. \\

We then define the Hubble expansion rate by
\begin{equation}\label{e.defH}
H=\frac{1}{3}\Theta.
\end{equation}
$H^{-1}$ characterizes the typical time scale of the expansion of the spacetime, or equivalently
\begin{equation}
R_H\equiv H^{-1}
\end{equation}
is the Hubble radius, that is the radius of the Hubble sphere (which is defined by this relation). It is not directly related to any causal limits (the speed of light does not enter its definition); but it is important in the discussion of the dynamics of perturbations of a FL universe, because it allows one to state for a mode of wavelength $\lambda$ whether or not the expansion of the spacetime affects its evolution; see  Eq.~(\ref{eq.X}) below. As a consequence, it plays an important role in structure formation in cosmology~\cite{PetUza13,MukFelBra80}.
 
\vskip0.25cm
\noindent[B] {\it\bf Apparent horizon (AH)}. 

Given bundles of outgoing and ingoing null geodesics of tangent vector $k^\mu$ satisfying Eq.~(\ref{e.null}) and $n^\mu$ similarly defined, one can similarly to the case of a fluid define the expansion of the bundles as
\begin{equation}
\hat \theta_+\equiv\frac{1}{2}\nabla_\alpha k^\alpha,\,\,
\hat \theta_-\equiv\frac{1}{2}\nabla_\alpha n^\alpha.
\end{equation}

Apparent horizons separate regions where either future or past null geodesics going either both inwards or both outwards converge, from where this is not the case. 

We distinguish outer (resp. inner) apparent horizons, AH($+$) (resp. AH($-$)) for outgoing (resp. ingoing) bundles. Each can occur in a future or past direction of time.

Future outer apparent horizons FAH($+$) are then defined to be spacelike 2-spheres where the local divergence rate $\theta_+$ for outgoing null geodesics vanishes after some initial 2-surface but not before. This implies that if the null energy condition $G_{\mu\nu}k^\mu k^\nu\geq0$ is satisfied, the outgoing as well as the ingoing null rays from this 2-sphere are from then on refocusing, and this will lead to conjugate points and  self intersections of these rays in the future \cite{HawEll73,Pen65}. This is the black hole case, and the case of recollapsing cosmologies in the future. Past outer apparent horizons PAH($+$) are defined to be spacelike 2-spheres where the local outgoing null ray divergence rate $\theta_+$ vanishes before some initial 2-surface but not after, leading to conjugate points in the past if the energy conditions are satisfied. This is the case in all realistic cosmological models.

Similarly there are Future inner apparent horizons FAH($-$) in the cosmological case where there is a (positive) cosmological constant.  They are then defined to be spacelike 2-spheres where the local divergence rate $\theta_+$ for ingoing null geodesics vanishes after some initial 2-surface but not before. This implies that if the null energy condition $G_{\mu\nu}k^\mu k^\nu\geq0$ is not satisfied, the outgoing as well as the ingoing null rays from this 2-sphere are from then on both diverging. This is the expanding de Sitter case. Past inner apparent horizons PAH($-$) are defined similarly. This is the collapsing de Sitter case.\\

Apparent horizons are not directly related to any causal limits, as their definition does not involve light cones, but are important in discussions concerning the existence of singularities if energy conditions hold, because the existence of conjugate points relates to limits on the causal future or past via the field equations and energy conditions \cite{HawEll73,Pen65}.  In general, the 3-dimensional apparent horizon surface made up of all 2-dimensional apparent horizons can be timelike, spacelike, or null, depending on whether matter is crossing the horizon or not (they will be null in the vacuum case). In the null case they will be related to event horizons.

Outer apparent horizons occur in a past directed way in cosmology, showing singularities should have existed in the past. There, they are not related to causal limits but are related to minima of apparent sizes of rigid objects (which occur for example at  a redshift of  $z=1.25$ in an Einstein-de Sitter universe) \cite{ell71,EllRot93}.
They occur in a future  directed way in astrophysical black holes, showing singularities should exist in the future \cite{HawEll73,Pen65}. There, they are related to causal limits in the future direction of time because a collapsing fluid leading to an astrophysical black hole results in the  existence of future directed apparent horizons that are associated with event horizons and production of Hawking radiation. Inner apparent horizons occur in a future directed way in de Sitter like phases of the universe, where they may be related to production of Hawking radiation (but which has no observable effects in a finite time \cite{jav}).

\section{Conformal diagrams}\label{sec1b}

The study of the global structures of a spacetime is simplified by the use of a representation introduced by Penrose. It is based on the idea of constructing, for any manifold $\mathcal{M}$ with metric $g_{\mu\nu}$, another manifold $\tilde{\mathcal{M}}$ with a boundary $\mathcal{J}$ and metric $\tilde g_{\mu\nu}=W g_{\mu\nu}$ such that $\mathcal{M}$ is conformal to the interior of $\tilde{\mathcal{M}}$, and so that the ``infinity'' of $\mathcal{M}$ is represented by the ``finite'' hypersurface $\mathcal{J}$. The last property implies that $W$ vanishes on $\mathcal{J}$. All asymptotic properties of $\mathcal{M}$ can be investigated by studying $\mathcal{J}$ (see Ref.~\cite{3.penrose} and Ref.~\cite{wini} for a pedagogical
intoduction).

\subsection{Construction of a Penrose diagram}

As a simple example, the Minkowki metric $\dd s^2=-\dd t^2+\dd r^2+r^2\dd\Omega^2$ can be written in terms of the advanced and retarded null coordinates,$v=t+r$ and $u=t-r$, as $\dd s^2 = -\dd v\dd u +\frac{1}{4}(v-u)^2\dd\Omega^2$ where $v\geq u$ and $v$ and $u$ range from $-\infty$ to $+\infty$. Then, one can compactify $u$ and $v$ by defining new rescaled null coordinates $\tan V= v, \qquad \tan U = u$ so that $-\pi/2< U\leq V <\pi/2$. Then, introducing the Minkowski like coordinates $T$ and $R$ by
$T=U+V$ and $R=V-U$ (such that $-\pi<T+R<\pi$ and $-\pi<T-R<\pi$, $R\geq0$) the Minkowski metric turns out to be conformal to the metric $\bar g$ given by $\dd\bar s^2 = -\dd T^2 + \dd R^2 +\sin^2R\,\dd\Omega^2$,  with conformal factor $W=\lbrace2\sin[(R+T)/2]\sin[(T-R)/2]\rbrace^{-2}$. Minkowski spacetime is thus conformal to a portion of the Einstein static spacetime (a cylinder $S^3\times \mathbb{R}$ which represents a static spacetime with spherical spatial sections). The boundary of this region therefore represents the conformal structure of infinity of the Minkowski spacetime.

This boundary can be decomposed into 
\begin{itemize}
\item two 3-dimensional null hypersurfaces, ${\mathcal J}^+$ and ${\mathcal J}^-$ defined by ${\mathcal J}^+=\left\lbrace V=\frac{\pi}{2},\quad |U|<\frac{\pi}{2}\right\rbrace$ and  ${\mathcal J}^-=\left\lbrace U=\frac{\pi}{2},\quad |V|<\frac{\pi}{2}\right\rbrace$ or equivalently by $T=\pm(\pi-R)$ with $R\in[0,\pi]$. The image of a null geodesic originates on ${\mathcal J}^-$ and terminates at ${\mathcal J}^+$, which represent future and past {\it null-infinity};
\item  two points $i^{+}$ and $i^{-}$ defined by ${i}^{\pm}:\quad U=V=\pm\frac{\pi}{2}$, or equivalently by $R=0$ and $T=\pm\pi$. The image of a timelike geodesic originates at $i^-$ and terminates at $i^+$. They represent future and past {\it timelike infinity}, that is respectively the start- and end-point of all timelike geodesics.
\item one point $i^0$ defined by ${i}^0:\quad U=-V=-\frac{\pi}{2}$,  or equivalently by $R=\pi$ and $T=0$. It is the start- and end-point of all spacelike geodesics so that it represents {\it spatial infinity}.
\end{itemize}
The fact that $i^\pm$ and $i^0$ are single points follows from the fact that $\sin R=0$. These are coordinate singularities of the same type as the one encountered at the origin of polar coordinates. The manifold $\tilde{\mathcal{M}}$ is regular at these points. $\mathcal{J}^-$ is a future null cone with vertex $i^-$ and it refocuses to a point $i^0$ which is spatially diametrically opposite to $i^-$. The future null cone of $i^0$ is $\mathcal{J}^+$ which refocuses at $i^+$.

Note that the boundary is determined by the spacetime and is unique but that the conformal extension spacetime (here the Einstein static spacetime) is not fixed by the original metric and is not unique since another conformal transformation could have been chosen.

\begin{figure}[t!]
\begin{center}
\resizebox{5cm}{!}{\includegraphics[clip=true]{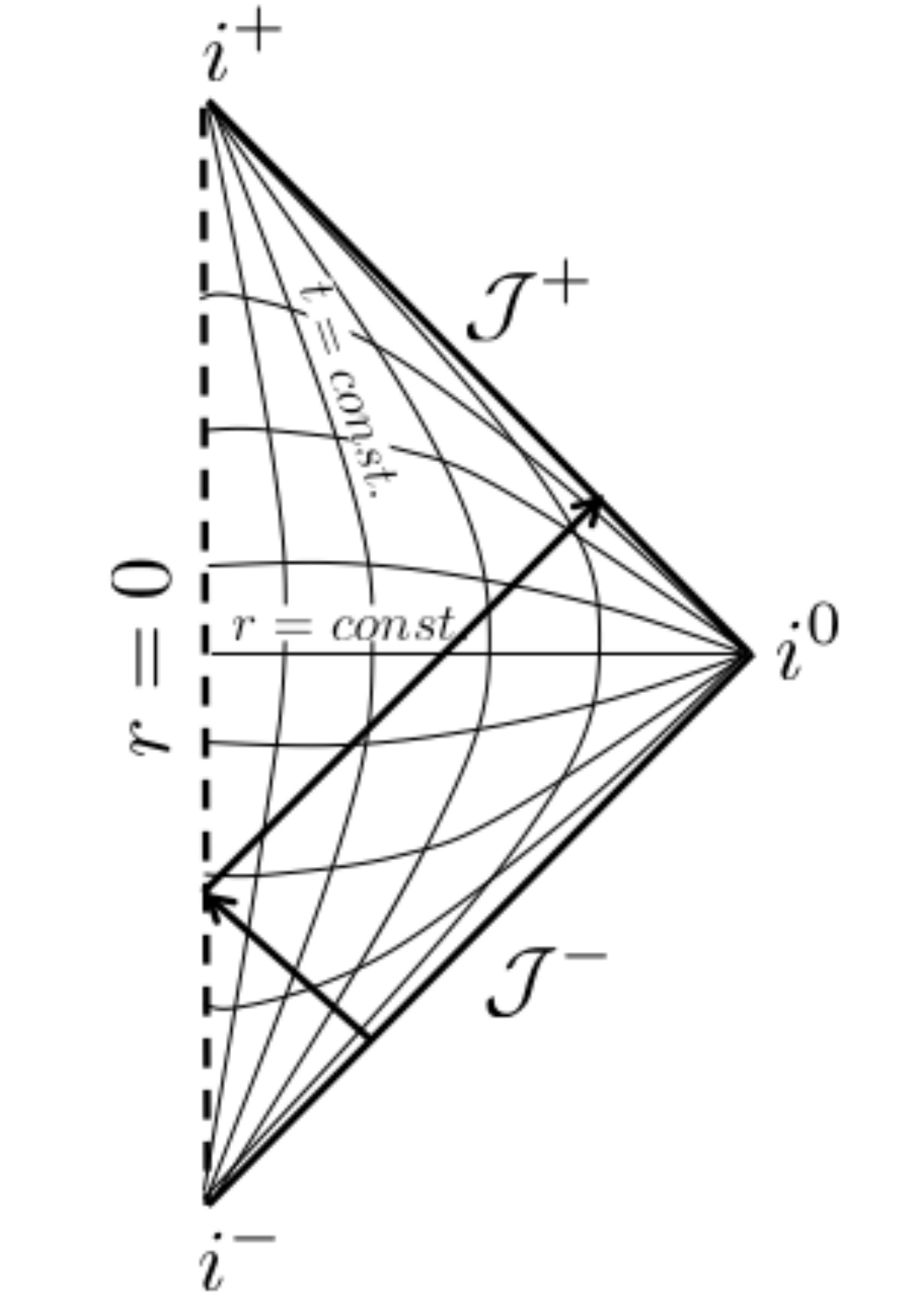}}
\end{center}
 \caption{\footnotesize Conformal diagram of Minkowski spacetime.}
  \label{fig2b}
\end{figure}

For any spherically symmetric spacetime, the Penrose diagram can be represented in the $(T,R)$ plane by ignoring the angular coordinates so that each point represents a sphere $S^2$. In this representation the Minkowski spacetime is a square lozenge (see Fig.~\ref{fig2b}). Null  radial geodesics are represented by straight lines at $\pm45\deg$ running from ${\mathcal J}^-$ to ${\mathcal J}^+$,  one of which apparently bounces when it passes through coordinate singularity at $r = 0$.

\subsection{Relation with the existence of horizons}

For most physically interesting spacetimes, $\mathcal{J}^+$ and $\mathcal{J}^-$ are either spacelike or null, which is closely related to the existence of one of the two types of non-local horizons described above.

If $\mathcal{J}^-$ is spacelike, the worldlines of the fundamental observers do not all meet on $\mathcal{J}^-$ at the same point. For any particular observer and event on its worldline close to $\mathcal{J}^-$, the past light-cone of this event will not intercept all the particles in the universe before it reaches $\mathcal{J}^-$ and there will be a particle horizon. If $\mathcal{J}^-$ is null, it is expected that all the worldlines of the fundamental observers pass through the vertex $i^-$ so that the past light-cone of a point $P$ will intercept all the worldlines and there is no particle horizon.

If $\mathcal{J}^+$ is spacelike, the worldline of any fundamental terminates on a point $O$ of $\mathcal{J}^+$ and the past light-cone of $O$ divides the universe into events that can be seen by the observer and events that he can never see; there is an event horizon. If $\mathcal{J}^+$ is null, all the worldlines will pass through the vertex $i^+$ so that $O=i^+$ and its past light-cone is $\mathcal{J}^+$ and there is no event horizon.

In conclusion,
\begin{eqnarray}
 && \mathcal{J}^-\,\,\hbox{spacelike}\,\,\Longleftrightarrow\,\,{\rm existence\,\, of\,\, a\,\,
 PH},\nonumber\\
 && \mathcal{J}^+\,\,\hbox{spacelike}\,\,\Longleftrightarrow\,{\rm existence\,\, of\,\, an\,\,
 EH}\nonumber.
\end{eqnarray}
We see from Fig.3 that neither horizon exists in Min-kowski spacetime for the usual static observers (with wordlines $r = const$).

\section{Causal structures in cosmology}\label{sec3}

\subsection{Geometry}\label{sec31}

In the standard cosmological model~\cite{HawEll73,PetUza13}, the universe is described by a Friedmann-Lema\^{\i}tre (FL) spa\-cetime with geometry
\begin{equation}\label{e.FL}
 \dd s^2 = -\dd t^2 + a^2(t)\gamma_{ij}(x^k)\dd x^i\dd x^j\ ,
\end{equation}
$t$ being the cosmic time (i.e. the proper time of fundamental observers), $a(t)$ the scale factor and $\gamma_{ij}$ the metric of constant time hypersurfaces, $\Sigma_t$, in comoving coordinates (Latin indices $i,j,\ldots$ run from 1 to 3). The matter 4-velocity is $u^\mu = \delta^\mu_0$. The conformal time $\eta$, defined by $\dd\eta =\dd t/a(t)$  can also be introduced to rewrite the metric  in the form
\begin{equation}\label{e.FL2}
 \dd s^2 = a^2(\eta)\left(-\dd\eta^2+\gamma_{ij}\dd x^i\dd
 x^j\right)\ 
\end{equation}
with $\gamma_{ij}\dd x^i\dd x^j=\dd\chi^2+f_K^2(\chi)\dd\Omega^2$ in spherical comoving coordinates so that $\chi$ is the comoving radial distance.  Here $K$ is the normalized spatial curvature  and respectively for $K=+1,0,-1$, $f_K(\chi) =(\sin(\sqrt{K}\chi/\sqrt{K},$ $\chi, $\- $\sinh(\sqrt{-K}\chi/\sqrt{-K})$ .

The geodesic equation~(\ref{e.null}) can be solved for radial null geodesics to obtain $k^\mu = E(u^\mu + e^\mu)$ with  $\dot E/E=-H$ and $e^\mu$ a constant unit spatial vector.
It follows that $E=k_0/a$, $k_0$ being a constant. Thus the energy $E$, and hence the frequency $\nu\propto E$, of any photon varies as the inverse of the scale factor and the redshift is given by
\begin{equation}
1+z=\frac{a_0}{a}.
\end{equation}
This can be measured from the comparison of an observed spectrum to a laboratory spectrum, from which one can deduce how much the universe has expanded since the light was emitted. Indeed the comoving radial distance $\chi$ cannot be measured directly, but can be obtained from the redshift; however this requires knowledge of the expansion law of the Universe, that is $a(t)$.

The characteristic distance and time scales of any Friedmann-Lema\^{\i}tre universe are fixed by the value of the Hubble constant~(\ref{e.defH}) which is give  by $H=\dot a/a$. The order of magnitudes of the Hubble time and radius are obtained by expressing the current value of the Hubble parameter
as $H_0 = 100\,h\, \hbox{km}\cdot\hbox{s}^{-1}\cdot\hbox{Mpc}^{-1}$ with $h$ typically of the of order $0.7$ so that the present Hubble distance and time are 
\begin{eqnarray}
 D_{{\rm H}_0}&=&9.26\,h^{-1}\times10^{25}\,\hbox{m}\nonumber\\
 &\sim& 3000\,h^{-1}\,\hbox{Mpc} ,\\
 t_{{\rm H}_0}&=&9.78\,h^{-1}\times10^{9}\,\hbox{years}\ .
\end{eqnarray}

This expansion rate allows one to estimate the age of the universe. From $\dd t=\dd a/aH$
we find
\begin{equation}
t_0   = t_{{\rm H}_0}\int_0^\infty\frac{\dd z}{(1+z)\mathbb{E}(z)}
\end{equation}
with $\mathbb{E}\equiv H/H_0$ being a function determined by the Friedmann equation,
\begin{equation}
 \frac{H^2}{H_0^2}= \frac{8\pi G}{H_0^2}\sum_i\rho_i - \frac{K}{a^2H_0^2}+\frac{\Lambda}{3H_0^2}
\end{equation}
for a universe of spatial curvature $K$ containing $i$ fluids with density $\rho_i$ and a cosmological constant $\Lambda$. Introducing the normalized density parameters $\Omega_i=8\pi G\rho_i/{3H_0^2}$, $\Omega_\Lambda=\Lambda/3H_0^2$, $\Omega_K=-K/3a_0^2H_0^2$, this equation takes the form
\begin{equation}
 \frac{H^2}{H_0^2}= \sum_i\Omega_i (1+z)^{3(1+w_i)}+\Omega_K(1+z)^2+\Omega_\Lambda
\end{equation}
assuming that each fluid has an equation of state $P_i=w_i\rho_i$ with $w_i$ constant.

The comoving radial distance $\chi$ of an object with redshift $z_*$ that is observed by an observer located at $\chi=0$, is obtained by integrating along a radial null geodesic.  Radial null geodesics between $(0,t_0)$ and $(t_1,\chi)$ are given by $\theta, \phi$ constant and $\dd\chi=\dd t/a$, so
\begin{equation}\label{eq:chi}
 \chi(t_0,t_1) := \int^{t_1}_{t_0}\frac{\dd t}{a(t)}\quad\hbox{},
\end{equation}
so that $ \chi(t_0,t_1)=\eta_1-\eta_0$ in conformal time. Changing to redshift as a parameter, it follows that
\begin{equation}
a_0\chi(z_*) = D_{{\rm H}_0}\int_0^{z_*}\frac{\dd z}{\mathbb{E}(z)}\ .
\end{equation}

\subsection{Characterization of the different horizons}\label{sec32}

Using these relations, the former definitions allow us to define and compute the different horizons defined in Section~\ref{sec1}.\footnote{We will not deal with apparent horizons here as they are not significant for causal limits in cosmology.}

\vskip0.25cm
\noindent[A] {\it\bf Event horizons}. 

A necessary and sufficient condition for the existence of an event horizon is that the integral $\chi(t_0,t_1)$ 
is convergent as $ t \rightarrow \infty$ (or $t = t_+$ if it has a finite future).  Indeed, then at any time~$t_0$, there exists a worldline
\begin{equation}
\label{eq:ehFRW}
\chi=\chi_{_{\rm EH}}(t_0):=\chi(t_0,t_+ 
)=\int^{t_+} 
_{t_0} \frac{\dd t}{a(t)}\ ,
\end{equation}
such that a photon emitted at $t_0$ from $\chi=\chi_{_{\rm EH}}(t_0)$ towards the origin reaches $\chi=0$ at
$t_+=+\infty$ if the universe expands forever, or at a finite $t_+$ if the future is finite  ($k=+1$, or in big-rip scenarios where the upper bound of the integral is finite). Any photon emitted at $t_0$ for  $\chi>\chi_{_{\rm EH}}(t_0)$ never reaches the origin and any photon emitted at $t_0$ for  $\chi<\chi_{_{\rm EH}}(t_0)$ reaches the origin in a finite time. The Universe has an event horizon since at each time $t_0$,  only the events  with  $\chi\leq\chi_{_{\rm EH}}(t_0)$ will ever be accessible to the observer $O$ at any time in the future.

\vskip0.25cm
\noindent[B] {\it\bf Particle horizons}. 

A necessary and sufficient condition for the existence of a particle horizon is that the integral  
$\chi(t_1,t_0)$ is convergent as $t_1 \rightarrow 0$ or $t_1 \rightarrow - \infty$, depending on whether or not $a(t)$ continues for negative values of $t$. From now on we will consider the first case only; the other is similar.

Then at any time $t_0$, any particle such that the comoving radial coordinate  $\chi>\int_0^{t_0}a^{-1}\dd t$ has not yet been observed by an observer $O$ at the origin. The 2-dimensional surface $(t_0, \chi_{_{\rm PH}}(t_0), \theta, \phi)$ defined by
\begin{equation}\label{eq:phcosm}
\chi_{_{\rm PH}}(t_0):=\chi(0,t_0)=\int_0^{t_0}\frac{\dd t}{a(t)}
\end{equation}
defines the particle horizon at a given time $t_0$ and thus divides all particles into two sub-families: the ones that have been observed at $t_0$ or before $t_0$ [$\chi\leq\chi_{_{\rm PH}}(t_0)$] and the ones that have not yet been observed [$\chi>\chi_{_{\rm PH}}(t_0)$], i.e. they lie beyond the  particle horizon of $O$ at $t_0$. The particle horizon itself is the timelike 3-surface $\chi=\chi_{_{\rm PH}}(t_0)$ generated by all the fundamental worldlines through this 2-surface. It does not relate to particles moving away from us faster than the speed of light (see Ref.~\cite{EllRot93,vc} for a detailed discussion), despite the fact that many claim this to be so. The present physical size of the particle horizon is $d_{_{\rm PH}} = a(t_0) \chi_{_{\rm PH}}(t_0)$. Since $a(t)$ is a positive function, when $\chi_{_{\rm PH}}(t_0)$ exists (because  the integral in Eq.~(\ref{eq:phcosm}) converges), it is an increasing function of $t_0$ so that as time elapses, more and more particles are visible from $O$. Thus it is not possible for particles to enter the particle horizon and then leave it: once they are in causal contact in a FL spacetime, they are in causal contact forever. 

The hypersurface $\sigma:\{\chi=\chi(0,t)\}$ is the future lightcone emitted from the position of the observer at $t=0$ (the creation lightcone). The particle horizon at time $t_0$ can also be seen as the section at $t=t_0$ of this spacetime surface $\sigma$. In conformal time, the creation light cone  is a cone at $\pm \pi/4$ and the particle horizon at $t_0$ is a 2-sphere represented by a point where the creation light cone intersects $t=t_0$. 
\vskip0.25cm
\noindent[C] {\it\bf Primordial particle horizons}. 

Similarly the primordial particle horizon is defined as the particle horizon of an observer at $t_{\rm LSS}$, i.e.
\begin{equation}
\chi_{_{\rm PPH}}=\chi(t_{\rm LSS},0)=\int^{t_{\rm LSS}}_{0}\frac{\dd t}{a(t)},
\end{equation}
$t_{\rm LSS}$ being the time of recombination defined below. This limiting surface governs what causal interactions were possible up to $t_{\rm LSS}$, when the Cosmic Micro\-wave Background Radiation (CMB) was emitted; events at $t_{\rm LSS}$ whose comoving distance from each other exceeds  $ \chi_{_{\rm PPH}}$ cannot have influenced each other. 

\vskip0.25cm
\noindent[D] {\it\bf Visual horizons}. 

In the standard cosmological model, the universe remains transparent until the temperature of the photon bath has dropped enough to allow the formation of neutral hydrogen and helium. This happens at a redshift of order $z_{\rm LSS}\sim 1100$ which defines the time of last scattering, $t = t_{\rm LSS}$, which can be computed once the cosmological parameters are chosen. It means that the comoving visual horizon of an observer at $t_0$ (the worldlines through the 2-sphere $\Sigma_p$ mentioned above) is given by 
\begin{equation}\label{eq:VH11}
\chi_{_{\rm VH}}(t_0)=\chi(t_0,t_{\rm LSS})=\int_{t_{\rm LSS}}^{t_0}\frac{\dd t}{a(t)}.
\end{equation}
Clearly this is inside the particle horizon. 
One may notice here from the definitions that
\begin{equation}\label{eq:horizons_relate}
\chi_{_{\rm PH}}(t_0)=\chi_{_{\rm VH}}(t_0)+\chi_{_{\rm PPH}}.
\end{equation}

\subsection{Matter-radiation universes}\label{sec33}

\begin{figure}[t!]
\begin{center}
\hskip-1.75cm\resizebox{9cm}{!}{\includegraphics[clip=true]{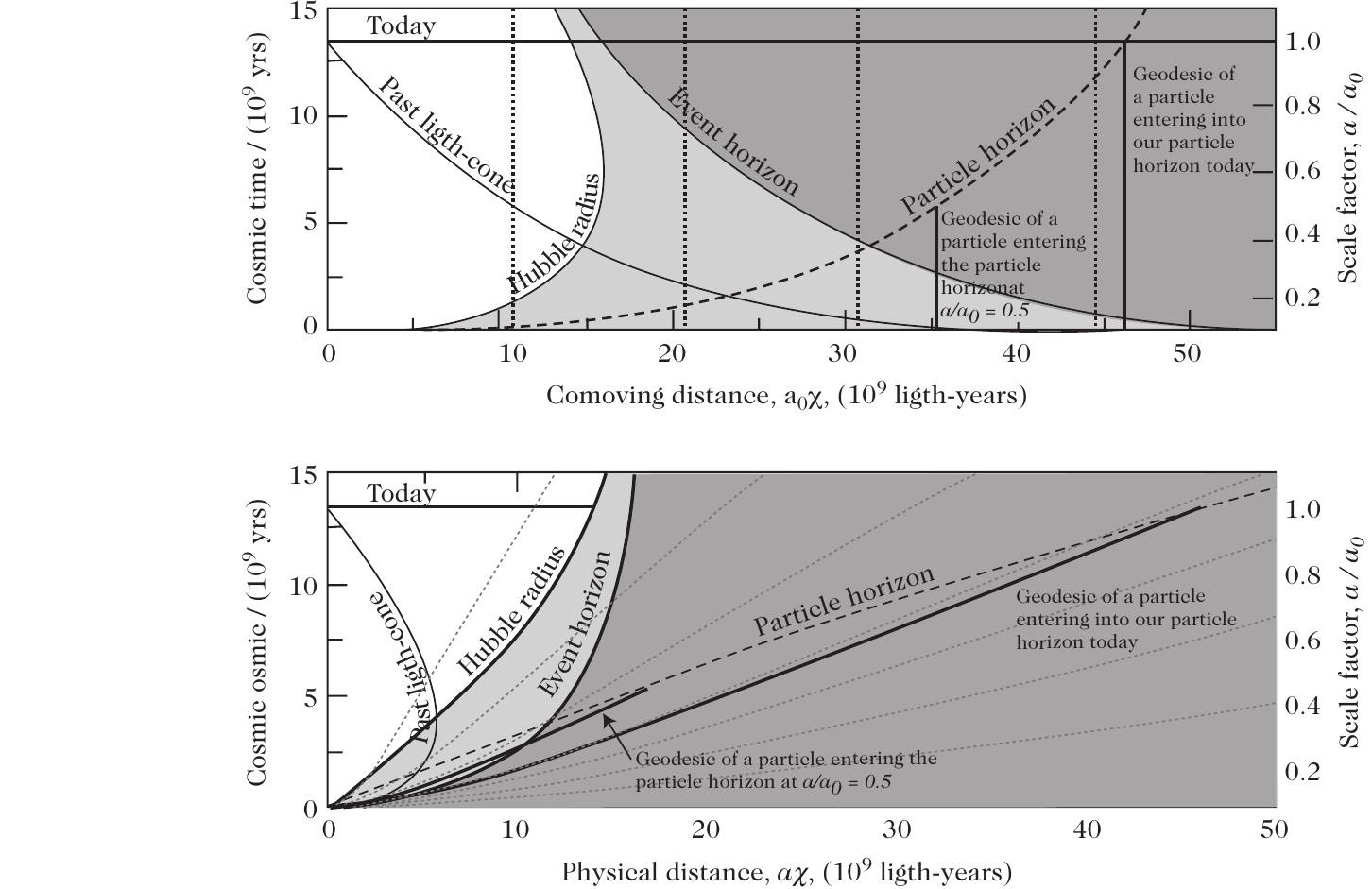}}
\end{center}
 \caption{\footnotesize The particle horizon at a given time is defined at different moments as the geodesic of the most distant observable 
 comoving particles at this moment (plain vertical lines). This surface can be visualized as the intersection of the creation light cone 
 with a 
 constant-time hypersurface. The equation of this surface is given by $\chi=\chi(0,t_0)$ 
 in comoving coordinates (top) and by $\chi=a(t_0)\chi(0,t_0)$ 
 in physical coordinates (bottom). This diagram represents a Universe with the following cosmological parameters 
 $(\Omega_m,\Omega_\Lambda)=(0.3, 0.7)$ and $h = 0.7$. From
Ref.~\cite{linedavis}}
  \label{fig3a}
\end{figure}

\begin{figure}[t!]
\begin{center}
\hskip-1.70cm\resizebox{8.7cm}{!}{\includegraphics[clip=true]{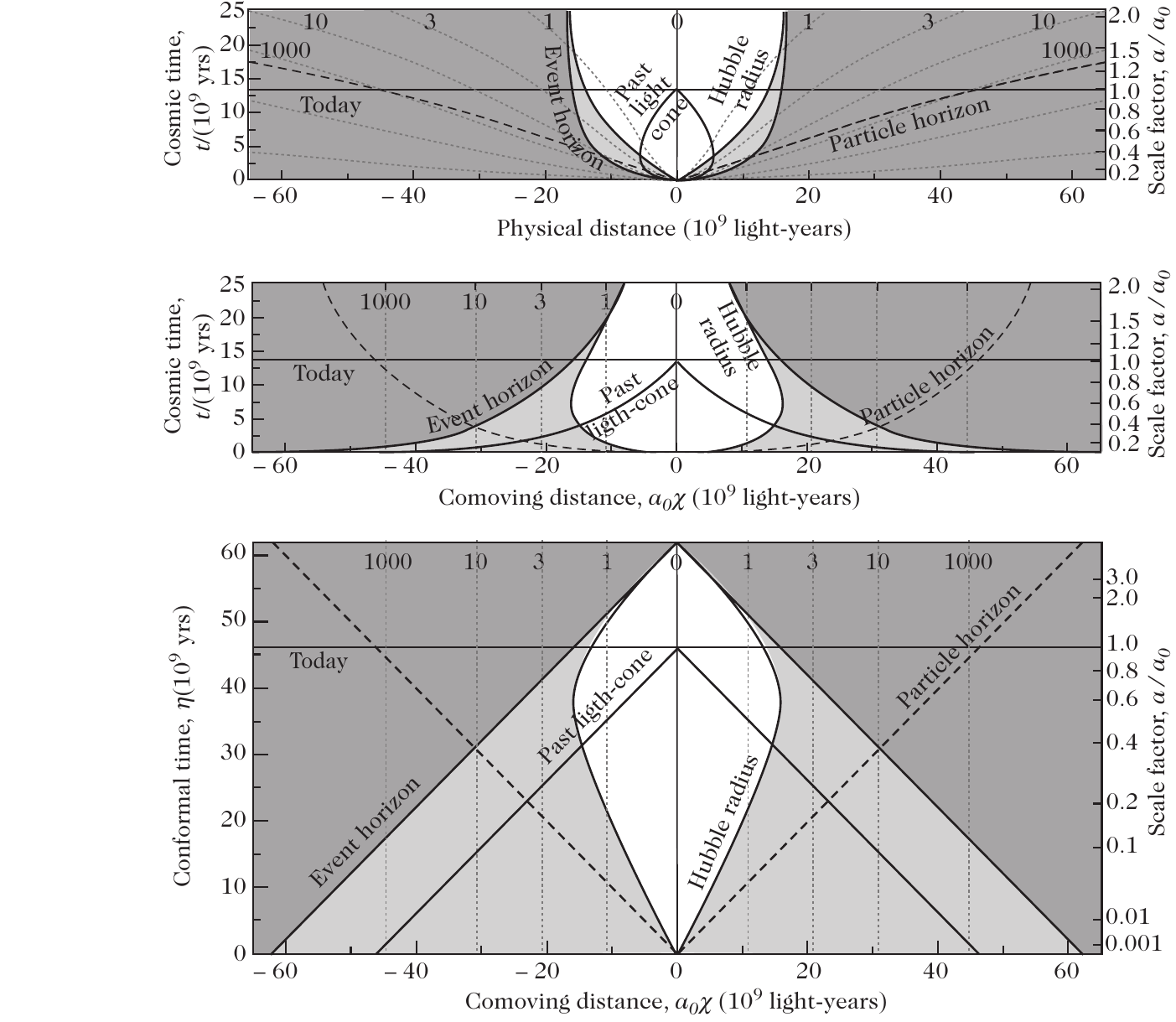}}
\end{center}
 \caption{\footnotesize Universe diagrams for a FriedmannÐLema\^{\i}tre space with parameters $\Omega_m = 0.3$, $\Omega_\Lambda = 0.7$ and $h = 0.7$. The first two diagrams represent, respectively, the physical distance, $a(t)\chi$, and the comoving distance, $\chi$, in terms of the cosmic time, $t$ (left scale) or in terms of the scale factor, normalized to 1 today (right scale). The last diagram represents the comoving distance in terms of the conformal time, $\eta$. The dotted lines represent the worldlines of comoving observers and our worldline is the central vertical line. Above each of these lines is indicated the redshift at which a galaxy on this worldline becomes visible for the central observer. We represent the past light cone for the present central observer and the event horizon corresponding to a similar light cone originated from timelike infinity (plain bold line). We also show the Hubble sphere (plain light line) and the particle horizon (dashed line). From Ref.~\cite{linedavis}.}
  \label{fig3b}
\end{figure}

Whenever the universe is filled by a fluid with a constant equation of state $w$, the scale factor behaves as $a\propto t^n$ with $3n=2/(1+w)$, provided $K=\Lambda=0$. The different horizons are then easily computed analytically. When $K\neq0$ or $\Lambda\neq0$ one can calculate them using the equations above; they are all depicted in Figs.~\ref{fig3a} and~\ref{fig3b}.

\vskip0.25cm
\noindent[A] {\it\bf Event horizons}. 

The integral $\int^\infty t^{-n}\dd t$ converges if and only if $n>1$.  Then, for the case of a single matter component 
with $K =\Lambda =0$, an event horizon exists if $w<-\frac{1}{3}$, i.e. if $\rho+3P<0$, that is if the strong energy condition is violated. So in general in this case for ordinary matter, {\em there is no event horizon.} However if $K
 = +1$ and the universe recollapses, or if $\Lambda>0$, there is indeed an event horizon for each observer. These two cases are different since for the former the bound on the integral is finite, which makes it converge, whereas for the latter, it is the growth of the scale factor that is sufficiently large to make the integral converge.

\vskip0.25cm
\noindent[B] {\it\bf Particle horizons}. 

The integral $\int_0 t^{-n}\dd t$ converges if and only if $n<1$, so there exists a particle horizon for a model with $K = \Lambda = 0$ if $w>-{1}/{3}$, that is if $\rho+3P>0$. 
There will also be particle horizons if $K>0$ or $\Lambda > 0$. 
Note that for a single fluid with a constant equation of state and $K=\Lambda = 0$, one can have either an event horizon or particle horizon, but not both at the same time. The two types of horizons are thus mutually exclusive in this particular case.

The physical diameter at a time $t_2$ for the particle horizon of an event that occurred at  $t_1<t_2$ is the limit
$$
 D_{\rm PH}(t_1,t_2) = a(t_2)\chi(0,t_1) . 
$$
It can be checked that if  $t_1$ and $t_2$ are two events from an era dominated by a fluid with constant equation of state $w$ and $K = \Lambda = 0$, then
\begin{equation}\label{eq.pph}
 D_{\rm PH}
 = \frac{6(1+w)}{1+3w} t_2\left[1-
 \left(\frac{t_1}{t_2}\right)^{\frac{1+3w}{3+3w}}\right].
\end{equation} 

\vskip0.25cm
\noindent[C] {\it\bf Visual horizon}. 
Unless we live in a small universe (we have seen right round the universe since last scattering because it has small enough closed spatial sections) there will always be a visual horizon because there are no divergent terms in  Eq.~(\ref{eq:VH11}). Indeed we have already seen back to the surface of last scattering, and (as shown by COBE and Planck observations) it covers the entire sky. That is the furthest  we will ever be able to see; and what we see is not infinite, as would be the case if we could see all the  matter in a $K=0$ universe with its standard topology .

\vskip0.25cm
\noindent[D] {\it\bf Hubble horizon}. 

The Hubble radius at time $t$ with $a\propto t^n$ is given by 
\begin{equation}
D_H(t) =n/t.
\end{equation} As pointed out above, this is locally defined and is not directly related to the past light cone or causal limits. This applies to the matter dominated period after recombination till quite recent times (when the universe started to accelerate), with $n = 2/3$.

If $t_1\ll t_2$ then the particle horizon diameter~(\ref{eq.pph}) is proportional to the Hubble radius at the time $t_2$
$$
 D_{\rm PH}(t_1,t_2) \simeq \frac{4}{1+3w}D_{\rm H}(t_2)\ .
$$
This is at the origin of the confusion between the Hubble radius and an horizon and the fact that one often assigns causality properties to the Hubble horizon. This property does not hold in other contexts such as during inflation.\\

\vskip0.25cm
\noindent[E] {\it\bf Penrose diagrams}. 

When written in terms of  conformal time, FL 
spacetimes with Euclidean spatial sections ($K=0$) are obviously conformal to Minkowski spacetimes. It follows that they map onto a part of a region representing Min\-kowski spacetime in the Einstein static universe. The actual region is determined by the range of variation of $\eta$. For $\Lambda=0$ and $P>0$, $0<\eta<\infty$ so that it is conformal to the upper half of the Minkowski diamond defined by $T>0$ with a singularity boundary, $T=0$. See Fig.~\ref{fig4}.

Friedmann-Lema\^{\i}tre spacetimes with spherical spatial sections ($K=+1$) are conformal to the Einstein static spacetime (with  the substitutions $\eta\rightarrow T$ and $\chi\rightarrow R$). They are thus mapped into the part of this spacetime determined by the allowed values for $\eta$. There are three general possibilities. When $\Lambda=0$, $\eta$ varies from 0 to $\pi$ if $P=0$ and from 0 to $\alpha<\pi$ when $P>0$. The Friedmann-Lema\^{\i}tre spacetime is thus conformal to a square region of the Einstein static spacetime so that both $\mathcal{J}^+$ and $\mathcal{J}^-$ are spacelike and represents two singularities. When $\Lambda\not=0$, $\eta$ can vary from $0$ to $\infty$ for the hesitating Universes  or from $-\infty$ to $\infty$ for the bouncing Universes. The FL 
spacetime is thus conformal to either the half of or the entire Einstein static spacetime. As in the case of the de~Sitter space (see Sec.~\ref{sec:desitter}), the conformal region will be a square of the Einstein static space and $\mathcal{J}^+$ and $\mathcal{J}^-$ are also spacelike but do not necessary represent a singularity.

In the case of Friedmann-Lema\^{\i}tre spacetimes wi\-th hyperbolic spatial sections ($K=-1$), the metric can be brought to its conformal form by means of the coordinate transformation
\begin{small}
\begin{eqnarray}
 T={\rm arctan}\left(\tanh\frac{\eta+\chi}{2}\right)+{\rm arctan}
 \left(\tanh\frac{\eta-\chi}{2}\right)
 \nonumber\\
 R= {\rm arctan}\left(\tanh\frac{\eta+\chi}{2}\right)-{\rm arctan}\left(\tanh\frac{\eta-\chi}{2}\right).
 \nonumber
\end{eqnarray}
\end{small}
Again the exact shape of this region depends on the matter content (equation of state and $\Lambda$).

We see from these examples that some parts of the boundary correspond to the big-bang singularity $a=0$. When $P> 0$ and $\Lambda\geq 0$, the initial singularity is spacelike, which corresponds to the existence of a particle horizon. If $K=+1$ or $\Lambda\geq 0$, the future boundary is spacelike, which signals the existence of event horizons for the fundamental observers

\begin{figure}[t!]    
\begin{center}\label{fig4}
\resizebox{7cm}{!}{\includegraphics[clip=true]{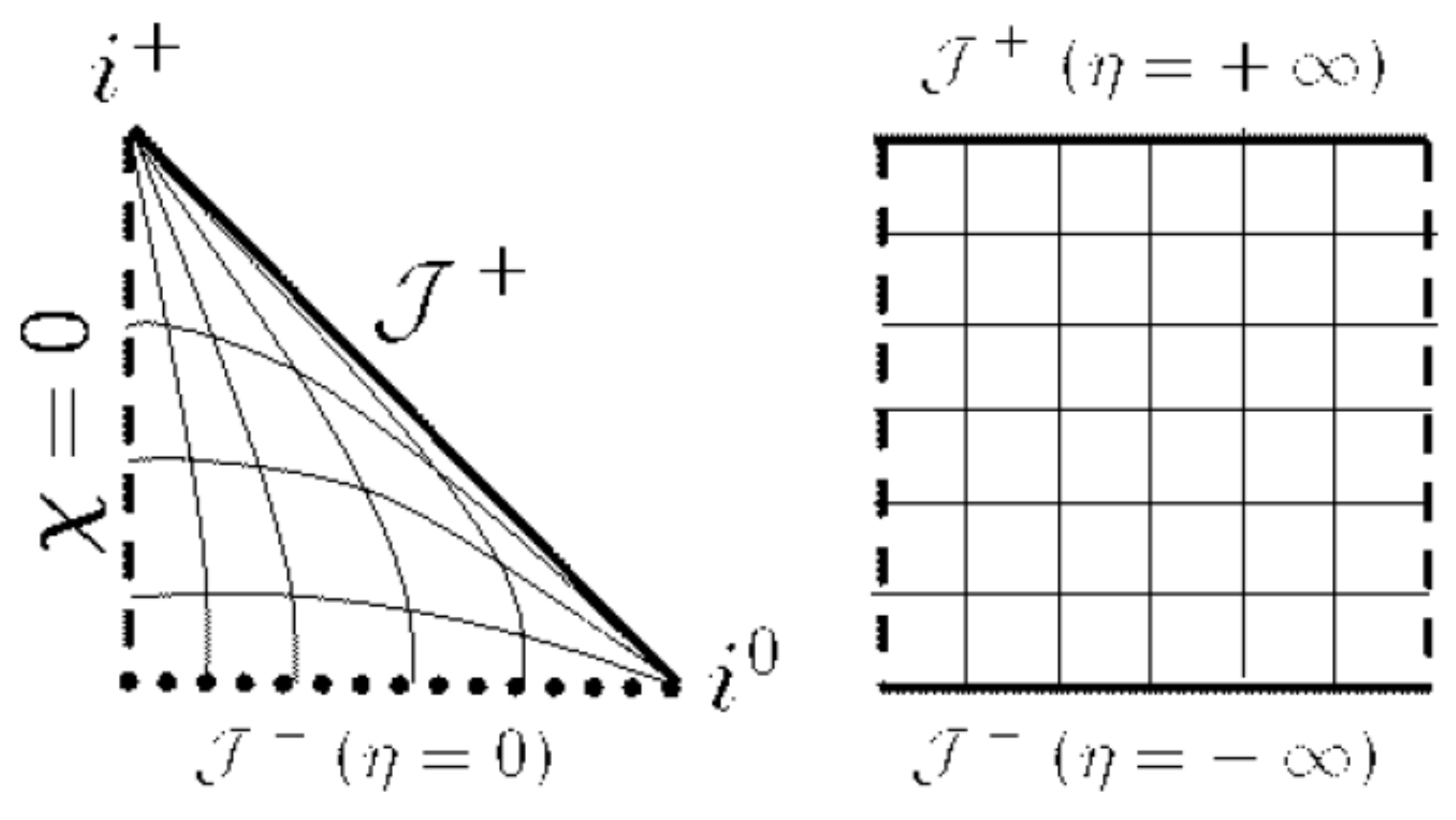}}
\end{center}
 \caption{\footnotesize Conformal diagram of the Friedmann-Lema\^{\i}tre spacetimes with Euclidean spatial sections with $\rho= 0$ and $P > $0 (left) and de Sitter space in the spherical slicing in which it is geodesically complete (right) -- dashed line corresponds to $\chi=0$ and $\chi=\pi$; see \S~\ref{sec33}[E].}
  \label{fig4}
\end{figure}

\subsection{Implication for the big-bang model}\label{sec2a}

In the standard big bang model, the universe is filled by a mixture of matter and radiation and a non vanishing cosmological constant. This means that (1) the universe was opaque before last scattering, so the visual horizon was determined by the time of last scattering,  and  (2) at late time the evolution of the scale factor is well approximated by
$$
a(t)=\left(\frac{1}{\Omega_{\Lambda0}}-1\right)^{1/3}
      \sinh^{2/3}\left(\frac{3\alpha t}{2}\right)\ ,
$$
where $\alpha=H_0\sqrt{\Omega_{\Lambda0}}$. 
Thus in the future our universe behaves as a de Sitter universe. As a consequence, unless it has a non-simply connected topology with small enough spatial sections,  we cannot see most of the matter in the universe. 

The expansion of the universe being accelerated, this also means that the worldlines of comoving structures are going out of our Hubble radius at the present time. Note that they remain in principle observable since they can never exit the visual and particle horizons once they have entered them. In practice, their apparent luminosity will drop very fast, making them fade out.

\subsection{Superhorizon modes}\label{sec2a2}

As far as cosmological perturbations are  concerned, the physical wavelength of any perturbation scales as $a(t)$. Since $\ddot a<0$ during a matter dominated or radiation dominated era, this  means that each physical wavelength will become smaller than the Hubble radius during such an era, if it lasts long enough. This leads to the notion of {\it super-Hubble} and {\it sub-Hubble} modes. Consider a mode with comoving wave-number $k$, it is said to be super-Hubble if its wavelength $\lambda\propto a/k$ is larger than Hubble radius $1/H$ and sub-Hubble otherwise. Thus,
\begin{eqnarray}
{\rm super-Hubble}: & k < aH, \nonumber\\
{\rm sub-Hubble}: & k > aH .\nonumber
\end{eqnarray}
As long as $a\propto t^n$, $a H\equiv{\cal H}$ is the comoving Hubble parameter and scales as $\eta$, so that this condition takes the form $k\eta<1$ or $k\eta>1$; this shows super-Hubble modes becomes sub-Hubble as $\eta$ increases. This distinction is important for the dynamics of perturbations. The relevant equations are partial differential equations involving only a Laplacian, so typically are of the form 
\begin{equation}\label{eq.X}
 \ddot X + H\dot X + k^2X=0
 \end{equation}
in Fourier space \cite{MukFelBra80,Muk05,PetUza13}. The term in $k^2X$ becomes important for sub-Hubble modes while it is negligible when the mode is super-Hubble. This transition, and when it takes place with respect to matter-radiation equality, has an importance for the growth of large scale structures \cite{PetUza13,MukFelBra80}.

However, historically, these modes have been cal\-led super-horizon and sub-horizon, the reason being that (as shown above) for a universe filled with matter and radiation, the Hubble radius and particle horizon are of the same order. Nevertheless this is confusing since causality (i.e. effects associated with the speed of light) is nowhere at work here. Besides this statement does not hold anymore during an inflationary or cosmological constant dominated era.

\subsection{The Horizon Problem} 
The CMB, originating at the surface $\Sigma_p$, delimits our visual horizon today. The horizon problem states that  in a matter and radiation dominated universe the visual horizon, which is of the order of the particle horizon, is much larger than the primordial particle horizon. This implies that the last scattering surface should be composed of many independent causally disconnected regions. One can then
not causally explain how these regions have thermalized for the CMB to enjoy a perfect black body spectrum with the same temperature over the whole sky. 

Let us 
emphasize that in cosmology one often refers to modes as {\it super-horizon} if they were {\it super-Hubble} at the time of last scattering. While these modes imprinted temperature fluctuations in the CMB, in particular giving rise to the Sachs-Wolfe plateau, they are {\it sub-horizon} at  at earlier times if they find their origin in an inflationary mechanism 
even if they were {\it super-Hubble} at last scattering. 
 Indeed one puzzle of the standard hot big bang model without inflation was to explain the existence of {\it super-Hubble} (and in that case also {\it super-horizon}) correlations at the time of last scattering.

\subsection{How far beyond the visual horizon can we extract information}\label{sec:limitsobs}

One can try to quantify the largest wavelength that can imprint a detectable signature on the CMB, i.e. to determine how far beyond the visual horizon one can observationally probe. 

An order of magnitude can be obtained by first considering the largest mode that needs to be included in a computation of the angular power spectrum, $C_\ell$. The CMB temperature anisotropy angular power spectrum is obtained as a convolution of the primordial power spectrum  $P(k)$, which is determined by the inflationary physics, and a transfer function $T_\ell(k)$, which is determined by the evolution of the post-inflationary perturbations: thus
\begin{equation}
 C_\ell=\frac{2}{\pi}\int P(k)T_\ell^2(k(\eta_0-\eta_{LSS}))\frac{dk}{k}.
\end{equation}
Formally, this integral runs from $k=0$ to $k=\infty$. In practice, it is computed with a lower cut-off $k_{\rm min}$ that has to be adjusted in such a way that it does not affect $C_\ell$. Given the shape of $T_\ell$ the main contribution to this integral is given by modes such that $k(\eta_0-\eta_{\rm LSS})\sim\ell$. What is the smallest $k_{\rm min}$ that needs to be considered? The answer is three-fold,  using the fact that in the standard concordance model $3\eta_0\sim a_0H_0$. 

(1) From a theoretical point of view, one needs to consider a $k_{\rm min}$ such that $k_{\rm min}(\eta_0 - \eta_{\rm LSS})$ is much smaller than the value at which the spherical Bessel function of order $\ell$ $j_\ell(x)$ peaks, i.e., $x\sim \ell$. Therefore, in practice it suffices to take $k_{\rm min} \eta_0 \ll 2$, the exact value mattering only for the quadrupole. This corresponds to
\begin{equation}
\lambda_{\rm max}({\rm pert})\sim200 R_{H_0}/a_0.
\end{equation} 
Low multipoles of the angular power spectrum (referred to as Sachs-Wolfe plateau) correspond to mo\-des that are super-Hubble at the time of last-scattering. The first acoustic peak, round  $\ell\sim220$ corresponds to the sound horizon at last-scattering~\cite{huphd}, which is a fraction of the Hubble radius at that time. 

(2) When taking into account the cosmic variance, which is large at small multipoles (typically below a $\ell$ of order 10), one can argue that modes larger than 
\begin{equation}
\lambda_{\rm max}({\rm cmb})\sim(10-20) R_{H_0}
\end{equation} do not leave significant signatures on the CMB. This explains also why a spatial topology with size smaller than the last-scattering diameter leads to a lack of power in the Sachs-Wolfe plateau (since the largest wavelength fixed by the size of the universe is smaller than $\lambda_{\rm max}$). This is also the case with polarisation and in particular $B$-modes~\cite{Lee:2014cya}. 

(3) Taking into account the observations, which have their own error bars and suffer from a galactic cut and many astrophysical effects, the question is when can one distinguish, given the same set of data, two models: one with a non-zero $k_{\rm min}$ and one with $k_{\rm min}=0$. This question can only be answered in a model dependent way. For instance, one can use a compact space and increase the size of its fundamental domain until it cannot be distinguished from an infinite space. In that particular case, topology induces multipole correlations, i.e. the correlator of the coefficients $a_{\ell m}$ of the expansion of the temperature field in spherical harmonics do not satisfy $\langle a_{\ell m}a_{\ell m}^*\rangle \propto C_\ell \delta_{\ell\ell'}\delta_{mm'}$. The use of the information encoded in the full correlation matrix (of the temperature and polarisation) has been investigated in the case of topology, using the notion of Kullback-Leibler distance, to conclude that at best one can probe topologies of size up to $1.15$ times the Hubble radius today (see Fig. 18 of Ref.~\cite{fabre}).  Thus 
\begin{equation}\label{eq:obslim}
\lambda_{\rm max}({\rm obs}) \sim(1.15) R_{H_0}/a_0
\end{equation} 
is the most that one can realistically probe by CMB observations. Note that this will be true whether or not inflation took place and that the numerical value may depend slightly on the spectral index. Also any observations of matter features, such as the Baryon Acoustic Oscillations, will probe scales less than this, because the corresponding angular scales lie well within the visual horizon (the BAO corresponds to the first acoustic peak in the CMB angular power spectrum peak at about $1^o$, while the visual horizon size is $180^o$).  

Let us also emphasize that upper bounds on $\Omega_K$, which correspond to lower bounds on the curvature radius of the spatial sections, are often used to state that we are actually probing the universe on much larger scales than $\lambda_{\rm max}$. Such an argument assumes, however, both that topology remains trivial on scales much larger than the Hubble radius today, and the validity of the Copernican principle on scales much larger than the Hubble radius today. Neither of these assumptions can be tested, so such a claim does not rely simply on data, but on their extrapolation under hypotheses that can be neither checked nor falsified (actually the last one is false if we live in a chaotic inflationary universe).

\begin{figure}[t!]
\begin{center}
\resizebox{7.5cm}{!}{\includegraphics[clip=true]{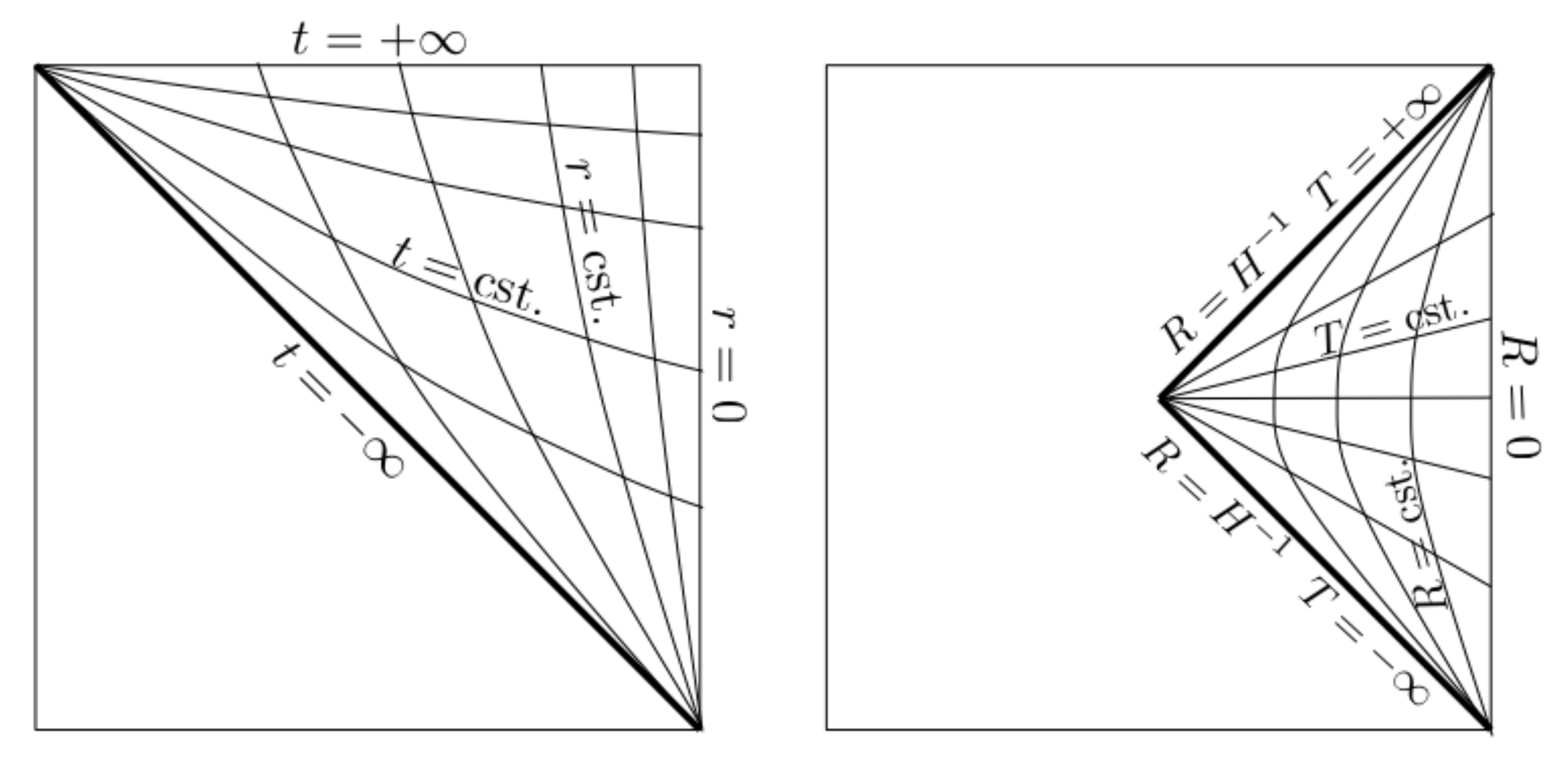}}
\end{center}
 \caption{\footnotesize Penrose diagrams of de Sitter space in the flat (left) and static (right) slicings that each cover only part of the whole de Sitter space, and that are both geodesically incomplete.}
  \label{fig4b}
\end{figure}

\section{Causal structures in inflationary models}\label{sec4}

During inflation in the very early universe, the expansion of the universe was accelerated. This acceleration phase helps solve various 
cosmological problems, in particular the horizon problem, and leads to a coherent theory of structure formation.

\subsection{Different approaches}\label{sec:diff}

Following Ref.~\cite{Lev}, there have been two main approaches to studying quantum effects during inflation. Both lead to a drastic modification of the Penrose diagram of the universe. 

The first approach, originating in the 60s~\cite{dsschool}, is fairly often used in the superstring community, in particular in the context of holography and the thermodynamics associated with horizons and the so-called ``hot tin can" picture \cite{susskind03}. It uses the static form of the de Sitter metric (see Fig.~\ref{fig4b}) so that an observer at the origin would detect thermal radiation from $R=1/H$ with a temperature $T=H/2\pi$, which corresponds to vacuum polarization of the de Sitter geometry.

\begin{figure}[t!]
 \vskip-1.5cm
\hskip-.75cm\resizebox{9cm}{!}{\includegraphics[clip=true]{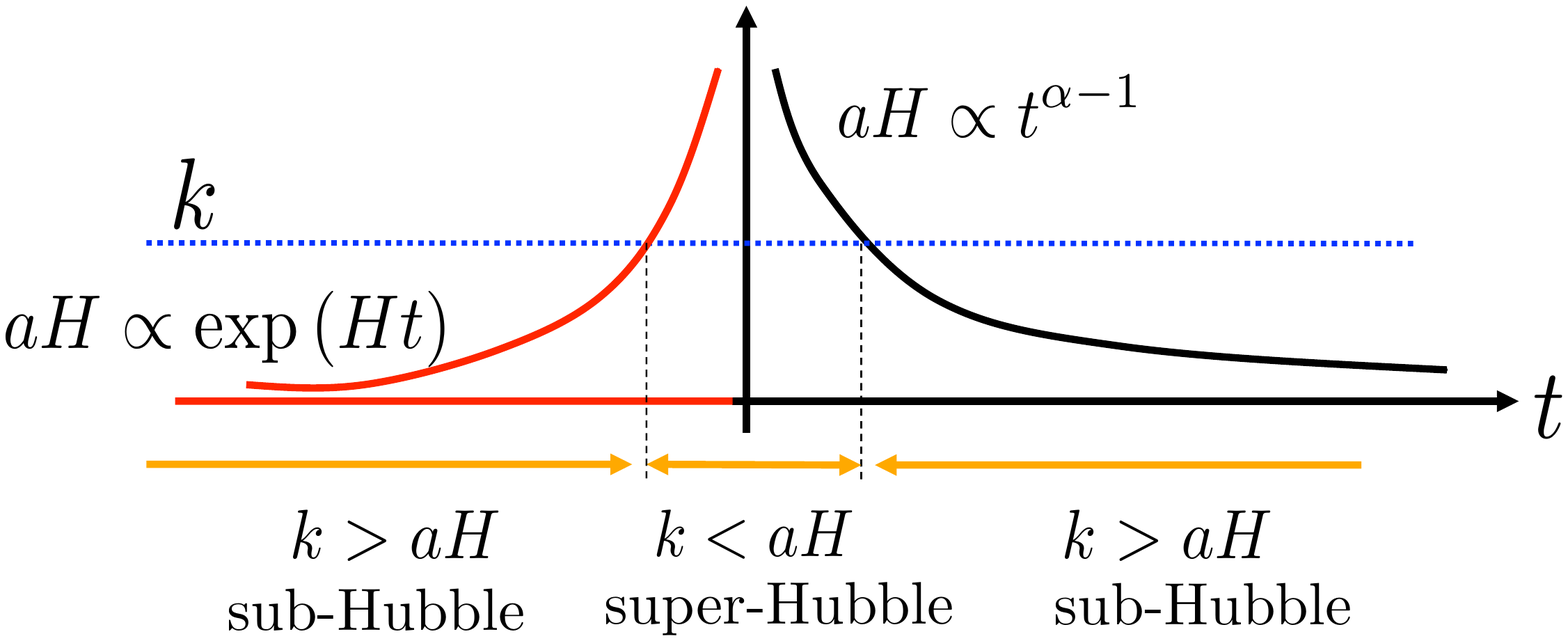}}
 \vskip-1.75cm
 \caption{\footnotesize Evolution of a mode during inflation and the hot big bang phases. A mode can be initially sub-Hubble, becomes super-Hubble during inflation and then sub-Hubble again.}
  \label{fig8}
\end{figure}

The second approach~\cite{timedep} is based on quantization of a scalar field in a time-dependent background described by an (almost) de Sitter space. It thus uses the flat representation of the de Sitter space (see Fig.~\ref{fig4b}). It follows that the Hubble radius during inflation is almost constant, which means that the comoving Hubble radius shrinks. Then, contrary to standard cosmology, this means that a comoving mode $k$ will become super-Hubble during inflation if it were initially sub-Hubble. This is summarized in Fig.~\ref{fig8}. This property is very important in the mechanism of the generation of initial perturbations during inflation, and is the explanation of the inflationary universe solution of the 
horizon problem for cosmological perturbations.

\subsection{de Sitter universe}\label{sec:desitter}

The de Sitter spacetime is a maximally symmetric spacetime. It enjoys many slicings, only one of them being geodesically complete. These different representations, corresponding to different choices of the family of fundamental observers,  are  \cite{HawEll73}:
\begin{enumerate}
 \item the spherical slicing in which the metric has a FL form with $K=+1$ spatial sections and scale factor $a\propto\cosh Ht$ with $H=\sqrt{\Lambda/3}$ constant. It is the only geodesically complete representation;
 \item the flat slicing in which the metric has a FL form with Euclidean spatial section, in which case $a\propto \exp Ht$;
 \item the hyperbolic slicing in which the metric has a FL form with $K=-1$ spatial sections and scale factor $a\propto\sinh Ht$;
 \item the static slicing in which the metric takes the form $ds^2=-(1-H^2R^2)dT^2+dR^2/(1-H^2R^2)-r^2d\Omega^2$.
\end{enumerate}
In each of the last three cases, the coordinate patch used covers only part of the de Sitter hyperboloid. 

\subsection{Horizons in a de Sitter space}

\noindent[A] {\it\bf Event horizon}.  

Since
$$
\chi_0=\int^\infty_{t_0} \frac{\dd t}{\exp
Ht}=\frac{\hbox{e}^{-Ht_0}}{H}<\infty\ ,
$$
there exists an event horizon for each of forms (1)-(3) of the de Sitter metric. There also exists one for form (4) because this static case is essentially like the Minkowski space case discussed above in Section \ref{sec22}[D] (see Fig.\ref{fig1}). 

\vskip0.25cm
\noindent[B] {\it\bf Particle horizon}. 
Is there a particle horizon?  In case (1), the $K=+1$ frame, we have $a(t) = a_0 \cosh(Ht)$ and the integral Eq.(\ref{eq:phcosm}) converges as $t_0 \rightarrow -\infty$ (there is no beginning), and there is a particle horizon. Thus there are both particle and event horizons in this bouncing eternally inflating cosmology.  Since in cases (2), (3), and (4) the universe is null geodesically incomplete in the past (the coordinate system is not global), this is not really a sensible question to ask. Nevertheless the integral  diverges as $t_0 \rightarrow -\infty$ in case (2) (there is no beginning to the expansion), so there is no particle horizon in this case. In case (3) there is a start to the expansion at $t=0$ and the integral diverges there, so there is no particle horizon. If the de Sitter expansion phase of the universe was finite, starting at some time $t_i$, then the answer depends on what happened before. 

\vskip0.25cm
\noindent[C] {\it\bf Hubble horizon}. 

Since $H=\sqrt{\Lambda/3}$ is constant in case (2), the physical Hubble horizon remains constant and the comoving Hubble horizon shrinks as
$H\exp(-Ht)$. At late times, this means that comoving worldlines are going out of our Hubble sphere. The same will be true for cases (1) and (3) at late times; see Ref.~\cite{krauss}. 

\vskip0.25cm
\noindent[D] {\it\bf Penrose diagram}. 
An exact de Sitter space is maximally symmetric so that there is no natural slicing. The choice of a particular slicing may lead to the fact that only a part of the Penrose diagram~\ref{fig4} is covered in flat or static representations. Using the coordinates
\begin{equation}
 R=\chi,\qquad T=2\, {\rm arctan}\left(\hbox{e}^{Ht}\right),
\end{equation}
with $0<R<\pi$ and $0<T<\pi$, the de Sitter spacetime with $K=+1$ (case 1) is  conformal to a square region of the Einstein static space
with conformal factor $W=H^{-2}\cosh^2(Ht)$ (see Fig.~\ref{fig4}). We see that this conformal diagram has a spacelike infinity both for timelike and null geodesics. However case (2) ($K=0$) covers only a triangular half of this domain, and case (4) (static) covers only a triangular quarter of that domain (see Fig.~\ref{fig4b}). 

The existence of these different representations of the de Sitter space is central in most arguments in the multiverse discussion (see below). 

\subsection{Cosmological model with an early inflationary phase}

 When inflation is driven by a scalar field, its slow-roll defines a natural time direction and slicing, but the spacetime is then only almost de Sitter, and in particular is then no longer  maximally symmetric.

\begin{figure}[t!]
\begin{center}
\resizebox{7cm}{!}{\includegraphics[clip=true]{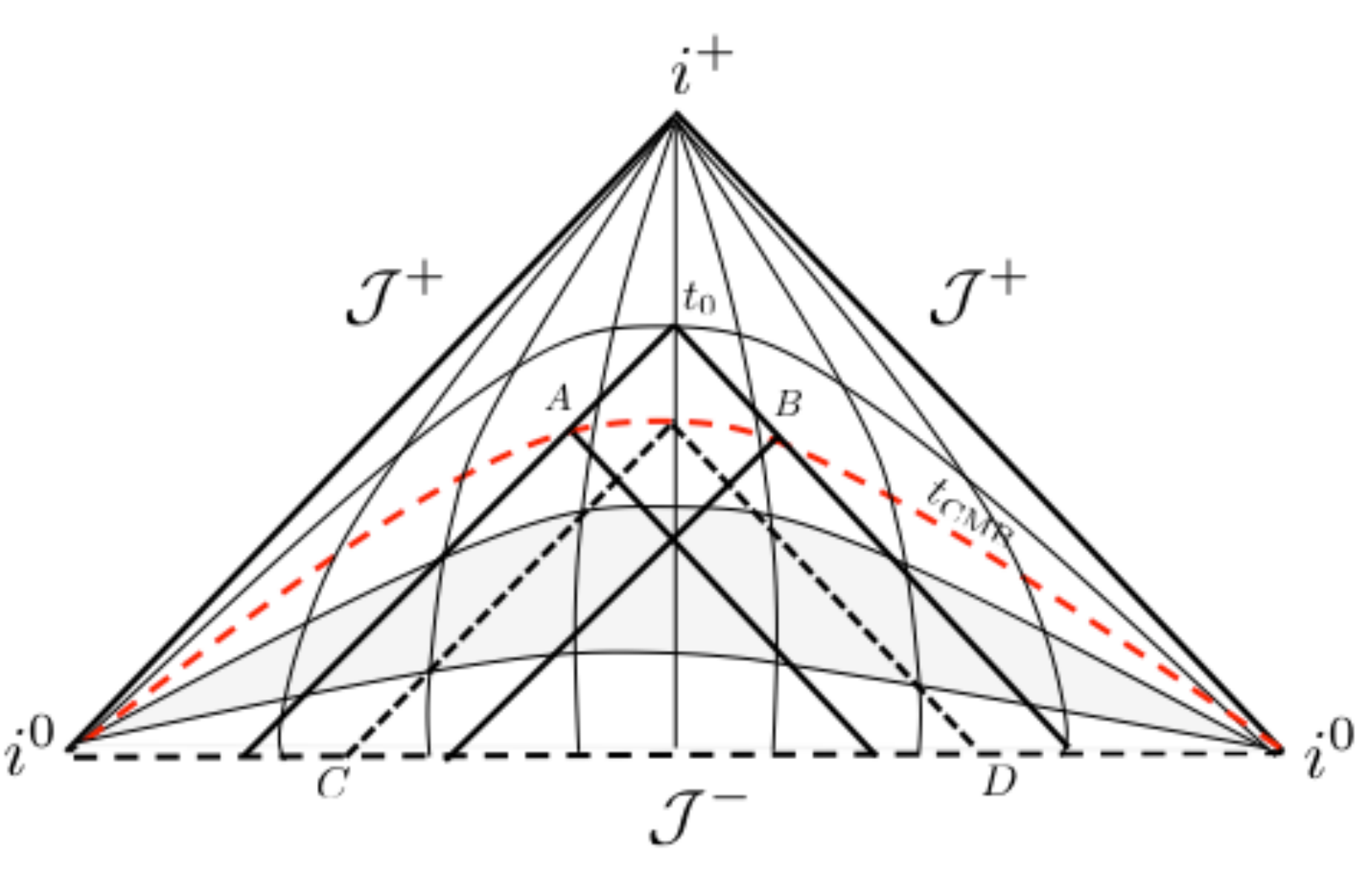}}
\end{center}
 \caption{\footnotesize Penrose diagram with an intermediate inflationary stage (shaded region). The dashed line respresent
 the last scattering hypersurface.  The light-cones from $A$ and $B$
 on the last-scattering hypersurface intersect only if there is a sufficiently long phase of inflation (hence solving the horizon problem).
 The dotted line represents the light-cone of an observer located on the last scattering surface. Hence $AB$ represents the Visual Horizon
 and $CD$ the Primordial Particle Horizon.}
  \label{fig6}
\end{figure}

The second consequence concerns the horizon pro\-blem. Consider that the universe underwent a phase of inflation from $t_i$ to $t_f$ in its early phase. Assuming it is spatially Euclidean with no cosmological constant, its Penrose diagram is similar to Fig.~\ref{fig4} (left). During the inflationary phase, the primordial particle horizon expands while the visual horizon remains unaffected. 
This means that, provided the inflationary phase lasts long enough, the visual horizon can become smaller than the primordial particle horizon (see Fig.~\ref{fig6}). Consequently the `horizon problem' goes away: there can have been time for causal influence to smooth the universe out on scales larger than the visual horizon \cite{PetUza13}. 

Finally, as noted above, comoving wavelengths will be leaving the Hubble horizon during this era, only to re-enter after the end of inflation. This plays a crucial role in structure formation \cite{PetUza13,MukFelBra80}.  

A key point is that inflation does not alter the limits on observations discussed in Section \ref{sec:limitsobs}, and particularly Eq.~(\ref{eq:obslim}) will still  hold.  There will exist super-Hubble perturbations \cite{Kol05} that can be probed by future galaxy surveys, but they will be seen to occur on angular scales (dependent on the relevant redshift range) that are smaller than the present day visual horizon, because the `Hubble Horizon' here relates to expansion rates in the past, not the present Hubble radius.\footnote{We thank Roy Maartens for comments on this topic.} Thus `super horizon modes' do not allow us to probe beyond the present Hubble scale. 

\section{Causal structures in alternative models}\label{sec6}

This section discusses the global structures of some alternative models used in the literature.

\begin{figure}[t!]
\begin{center}
\resizebox{7cm}{!}{\includegraphics[clip=true]{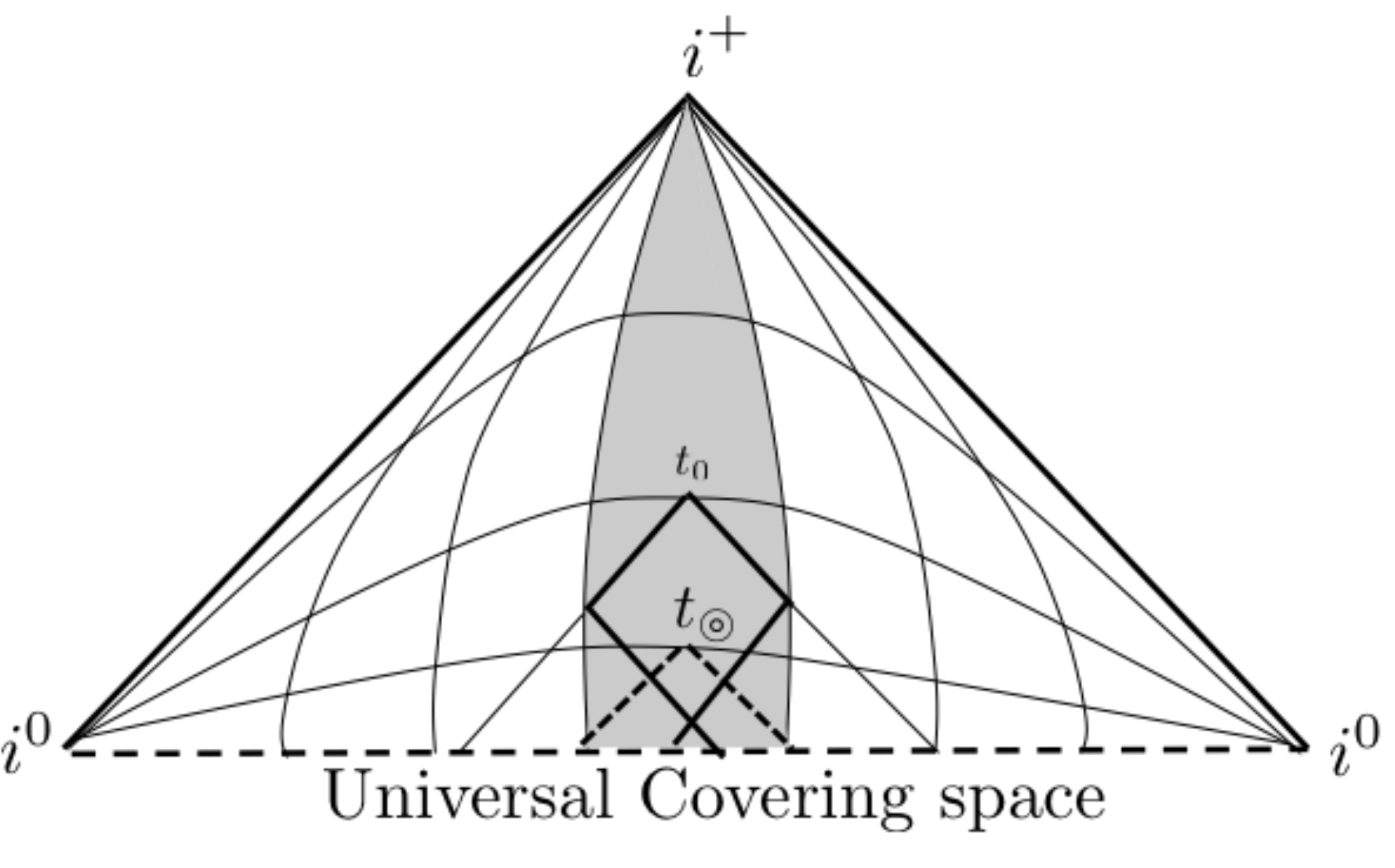}}
\end{center}
 \caption{\footnotesize Penrose diagram for a universe model with compact spatial sections. The shaded region corresponds to the fundamental polyhedron (i.e. the whole physical universe). The dashed line represents the light-cone from $t_{\circledcirc}$. For any $t_0>t_{\circledcirc}$ the light-cone warps around the universe so that there is no particle horizon.}
  \label{fig9}
\end{figure}

\subsection{Spatially compact universes}

A universe can enjoy compact spatial sections so that no (spatial) infinities occur and the volume is finite at each time. 
In standard cosmology where the spatial sections are homogeneous and isotropic Riemannian 3-manifolds, the topology can be described by its fundamental polyhedron, the faces of which are associated in pairs through the elements of a holonomy group. 
In that case new length scales enter the model and characterize the size and shape of the spatial sections. One often considers 
\begin{enumerate}
\item The (comoving) volume of the fundamental polyhedron,
\item The outside radius $r_+$, the radius of the smallest geodesic ball that contains the fundamental polyhedron, 
\item The inside radius $r_-$, the radius of the biggest geodesic ball contained in the fundamental polyhedron, and 
\item The injectivity radius $r_{\rm inj}$, half the length of the shortest closed geodesic. 
\end{enumerate}

 Depending on the expansion history of the universe, there may be 
a time $t_{\circledcirc}$ such that  
\begin{equation}
\chi(0,t_{\circledcirc}) \geq r_+,
\end{equation} 
with $\chi$ give by Eqn.(\ref{eq:chi}), and for $t \geq t_{\circledcirc}$ there will be no particle horizons: all matter  in the universe will be in causal contact, see Fig.~\ref{fig9}. If an inflationary epoch takes place, this can occur very early on (see e.g. Ref.~\cite{mixing} for an example). There will also then be no visual horizons \cite{topology}. The value of $t_{\circledcirc}$ depends on the topology and the matter content. For instance in a universe with $S_3$ spatial sections and no cosmological constant, $t_{\circledcirc}=t_{\rm crunch}$, the big-crunch time.  
,
\subsection{Chaotic inflation and multiverse}

\begin{figure}[t!]
\begin{center}
\resizebox{6cm}{!}{\includegraphics[clip=true]{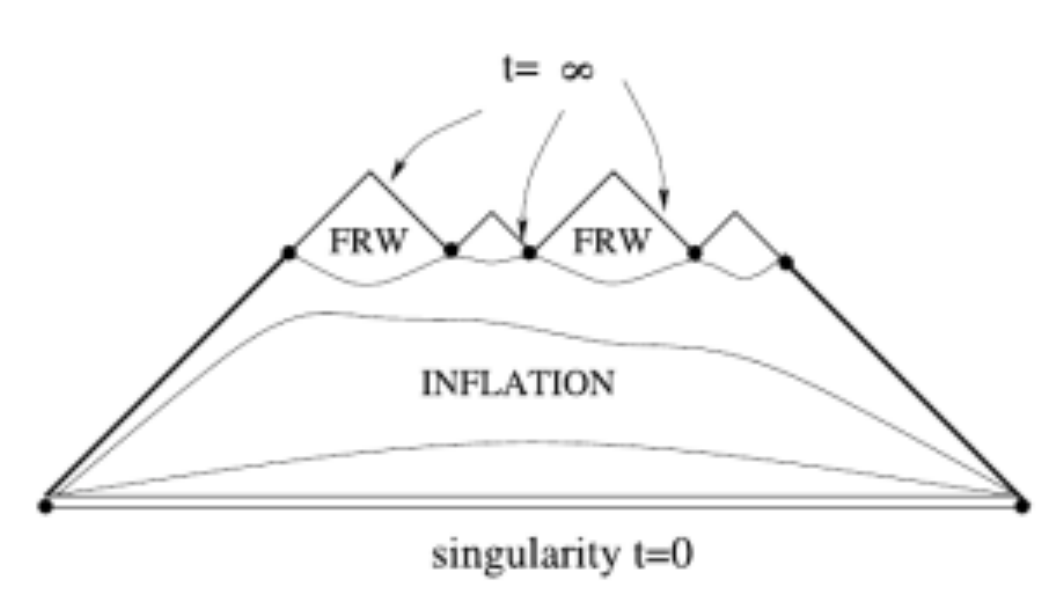}}
\end{center}
 \caption{\footnotesize Same as Fig.~\ref{fig6} with large scale quantum fluctuations. From Ref.~\cite{Lev}.}
  \label{fig7}
\end{figure}

In chaotic inflation, the value of the inflaton experiences 
large quantum fluctuations which result in spatial fluctuations of the number of $e$-folds of the inflationary phase. It follows that the universe is not homogeneous on very large scales. As the inflationary universe is not homogeneous globally, hypersurfaces of constant value of the inflaton field (which defined the onset and end of inflation) are no longer cosmic constant time hypersurfaces, so 
the end of the inflationary era corresponds to different times in different spatial domains. When the inflaton rolls beyond a critical value such that the expansion is no longer accelerated, it oscillates at the bottom of the universe potential and reheats a 
FL domain that is separated from the expansion of the other domains and has a $K=-1$ geometry. This happens repeatedly and thus forms a fractal distribution of FL patches, in the sense that ${\cal J}^+$ is a broken line (see Fig.~\ref{fig7}, also Ref.~\cite{susskind03}). Here 
 the relative sizes of the FL patches in the figure are irrelevant, because of the conformal freedom of the diagram. 
Note that the conformal structure of the future boundary is patchwise that of the left hand diagram in Fig.(\ref{fig4}) rather than the right hand one: that implies that $\Lambda$ is taken to be zero in each of these bubbles .

The implication of chaotic inflation on the global structure is three-fold: (1) global homogeneity is violated on large scales, (2) the asymptotic null future has a fractal structure, and (3) the universe has a large number of causally disconnected FL patches since decoupling. This is often called a multiverse because the different FL domains have different properties.

Note that if the cosmological constant $\Lambda$ is non-zero in any of these domains (as is probably the case in ours) then future infinity will be spacelike in that domain rather than null. The value of $\Lambda$ may vary across domains, or may be the same in all of them.
Usually the multiverse structure is invoked to explain the small positive value of $\Lambda$ we observe by having all possible values occurring in the various bubbles, and using anthropic selection effects to explain the small value we actually observe. In that case it is Figure~\ref{fig4} (right) that will represent the causal structure rather than Figure~\ref{fig7}, because $\Lambda$ will not be zero almost everywhere in the multiverse.\footnote{We assume here it only takes positive values. Things will be more complex if it also takes negative values.}

A key point is that all this structure still does not alter the limits on observations discussed in Section \ref{sec:limitsobs}, and 
particularly Eq.~(\ref{eq:obslim}) will still  hold.  One can note here that in a non-inflationary universe, the primordial 
particle horizon  (the particle horizon at recombination) has an angular size of $\theta_c \simeq 2.3^o$ (much less than the visual horizon 
size of $180^o$), so causal processes cannot lead to structure on larger scales in those universes \cite{Lee14}\footnote{It appears that what they call the `particle horizon at recombination' is what that horizon would be in the case where no inflation takes place. }. 
Inflation causes the  primordial particle horizon to become much larger than the visual horizon and so in principle allow observations to test these larger scales. However  as discussed above this will not occur for real observations, so Eq.~(\ref{eq:obslim}) is unaffected.

\subsection{Bubble collisions} 

In the multiverse description, one needs to carefully describe the process of nucleation in order to characterize the global structure. We cannot review all proposals but shall focus on the example of bubble nucleation in a de Sitter like background. 

One question is 
 whether neighbouring bubbles can leave an observational signature in our universe~\cite{freivogel,Freivogel:2014,Kleban}. They could potentially do so if bubble collisions take place. Whether such collisions will take place in a multiverse will depend on a competition between the expansion rate and the nucleation rate; there may be none or many, depending on how one chooses these parameters.   

\begin{figure}[t!]
\begin{center}
\resizebox{11cm}{!}{\includegraphics[clip=true]{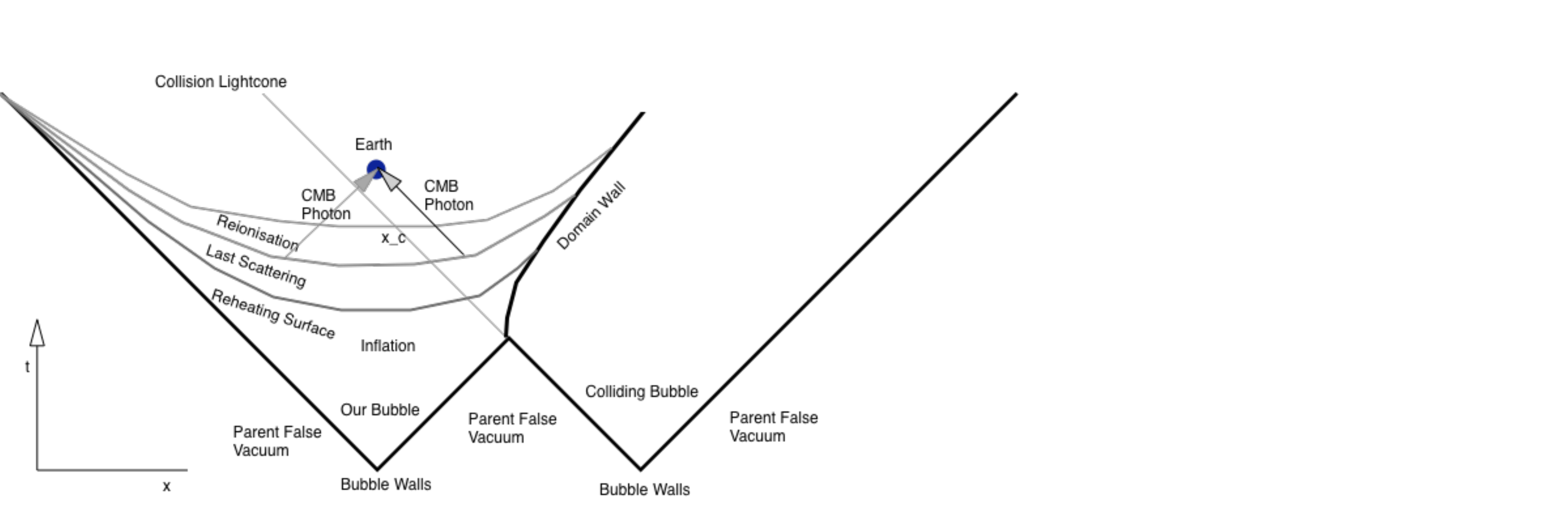}}
\end{center}
 \caption{\footnotesize Penrose diagram for a model with two colliding bubbles. The past light-cone of the observer contains part of the bubble wall (null) worldsheet so that the second bubble can influence the local physical conditions in our early universe, even though the bubble wall never enters our visual horizon: the nucleation event of the second bubble is outside our visual horizon but inside our Primordial Particle Horizon. From Ref.~\cite{Kleban}.}
  \label{fig8b}
\end{figure}

Figure~\ref{fig8b} presents the Penrose diagram of two colliding bubbles. In such a situation, the past light-cone of the observer contains part of the bubble wall worldsheet so that the second bubble can influence the local physical conditions in our early universe, even though the bubble wall never enters our visual horizon: the nucleation event of the second bubble is outside our visual horizon but inside our Primordial Particle Horizon. This means the collision could for example in principle  
lead to circles in the CMB sky as an observable effect~\cite{Kleban}.

In this model, the spatial sections have hyperbolic geometry and are often said to be infinite. Let us however emphasize that infinite spatial sections with $K=-1$ cannot in fact be instantaneously formed, as is often claimed in the literature. The reason is that point-like processes are required for this to happen; and such processes cannot occur when the nucleation process has a non-zero spatial extension~\cite{EllSto09}, which is the physically relevant case because of quantum effects. This is also implied by the structure shown in Figure \ref{fig8b} where these spatial surfaces run into a domain wall; what happens to them then depends on what is the other side of that wall. That would not be possible if infinite spatial sections had been instantaneously formed.  
 
Thus, if a multiverse forms by bubble nucleation in a de Sitter model in the $K=+1$ frame, it has to be spatially compact before the nucleation occurs and will remain so at any finite time after nucleation occurs (that is there will exist a foliation by compact spatial 3-surfaces, as in the de Sitter universe). Generically there will be particle horizons and event horizons in this case: indeed each bubble will have its own event horizon (if $\Lambda=0$ there), or there will be a set of event horizons for each observer (if $\Lambda \neq 0$ in the bubble, which is the more likely case).

The key point then is that the limits on observations discussed in Section \ref{sec:limitsobs}, and particularly Eq.~(\ref{eq:obslim}), still  hold {\em whatever the details of the nucleation process}. Observational access to other domains in a multiverse is highly restricted. It is possible we might get hints of existence of one or two other bubbles through CMB observations of bubble collision. However they will for example almost certainly not be able to show that the value of $\Lambda$ is different in the other bubble - which is what one would need to confirm the anthropic multiverse picture.    

\subsection{Static and dynamic pictures} 

Use of a mixture the different kinds of de Sitter patches (expanding volumes and static causal diamonds) in a multiverse picture in a complementary way, as sometimes proposed, is problematic. For example Susskind remarks in Ref.~\cite{susskind03} ``Pick a timelike observer who looks around and sees a static universe bounded by a horizon.'' He means an event horizon (such a static domain has no particle horizons for the static particles filling the patch), but you can't determine an event horizon in a finite time. So it is not in fact observable. 

But in any case the real universe is not static locally anywhere, because where it is not filled with matter, it is filled with radiation; and both are necessarily in a dynamic state. The dynamic local patches give a much better description. 

%
\subsection{Event Horizons}

Various writings claim to derive cosmologically relevant observational results that are related to the existence of event horizons in the multiverse context,  for example Ref.~\cite{Sus05}. The problem is that the event horizon has no link to present day observations, which are limited  by the visual horizon \cite{Ell06}. This suggestion does not relate to real observational cosmology.

As has been made clear above, event horizons only come into being in the far future of the universe or the multiverse, indeed as $t \rightarrow \infty$. Our past light cone, which is where we can carry out observations, lies well to the past of any event horizon that may exist. Consequently, existence of event horizons can have no effect whatever on any possible astronomical observation and so cannot have any relation to using observational data of any kind to constrain cosmological models. Furthermore the black hole information loss paradox and firewall issue are irrelevant to observational cosmology, because they are related to properties of event horizons.\\

One might reply that event horizons are used in some versions of structure formation calculations to deduce the existence of Gibbons-Hawking radiation in de Sitter spacetime, which is related to generation of quantum fluctuations (see the first option in section \ref{sec:diff}), and hence those event horizons do indeed have observational consequences through their effects on structure formation. The response is that such event horizons only exist if the de Sitter space is eternal; if the inflationary phase comes to an end at a finite time and thus leads to a standard big bang epoch, as in the standard model, there is no event horizon associated with the inflationary epoch because of that fact. (Hawking radiation may be associated with apparent horizons in the inflationary domain; but that will be redshifted so as to be negligible.)

The  essential nature of event horizons is related to the infinite future of an observer in a universe domain, not to anything that may have happened in the past or occurs in a finite time, and hence they are not related to structure formation, which can be properly determined from  the second option mentioned in Section \ref{sec:diff} (see Refs.~\cite{PetUza13,MukFelBra80}). And one does not need any concept of horizons in order to derive the quantum fluctuations that lead to structure formation in an inflationary universe: see e.g. Ref.~\cite{Muk05}.   
\\

In summary: event horizons play no role in observational cosmology, on a cosmological scale. Of course they may do so at an astrophysical scale when local black holes form: but that is a completely different story. 

\section{Discussion}\label{sec7}
Our argument can be summarised as follows:
\begin{enumerate}
\item Visual horizons are a key limit on what we can observationally test in a FL universe, and in a multiverse (if such exists). The Hubble ``horizon'' is not a causal limit but rather is a dynamical scale associated with structure formation in the expanding universe. It is the particle horizon that limits causal interactions up to the present day.

\item In an inflationary universe, the visual horizon can be much smaller than the particle horizon. ``Superhorizon modes'' can test conditions on scales greater than what the Hubble scale was at past times, but not greater than  roughly 1.15  times the Hubble scale/visual horizon at the present time (see Eq.~(\ref{eq:obslim})).

\item We can see the LSS where it intersects the visual horizon, apart from interference by intervening matter. We cannot see the LSS inside the visual horizon unless there are folds in our past light cone. But they are not likely to be large enough to be significant. However we can deduce some features of the interior of our past light cone through the SZ effect~\cite{SZeff}.

\item In a multiverse, unless we have had bubble collisions, all bubbles except our own are irrevocably inaccessible to all observational tests. But such collisions will only occur in a subset of multiverses; and if they do occur, they only give very limited access to data about a few bubbles. They will not give access to almost all bubbles (if a multiverse indeed exists).  

\item In a universe or multiverse, cosmological event horizons are irrelevant both to observational cosmology and to the origin of structure. The only observationally relevant apparent horizons are past directed and do not relate to causal limits other than those implied by the past null cone. They relate to minimal observed apparent sizes, associated with past directed closed trapped surfaces. 
\end{enumerate}

\Acknowledgements{We thank Thiago Pereira, Cyril Pitrou, and Alain Riazuelo. This work was supported by French state funds managed by the ANR within the Investissements d'Avenir programme under reference ANR--11--IDEX--0004--02, the Programme National Cosmologie et Galaxies, and the ANR THALES (ANR-10-BLAN-0507-01-02. JPU thanks Cape Town university for its hospitality during the early phase of this work and GFRE thanks the Institut d'Astrophysique de Paris for hospitality in the late phase of this work and the NRF (South Africa) for financial support.}


\end{document}